\documentclass[amsmath,trackchanges]{aastex702}
\usepackage{epstopdf}
\usepackage{graphicx}
\definecolor{red}{rgb}{1.0,0.0,0.0}

\newcommand{\Mj}[1]{$M_\mathrm{Jup}$}
\newcommand{\uas}{$\mu$as}
\newcommand{\kms}{km\,s$^{-1}$} 

\usepackage[utf8]{inputenc}

\begin{document}

\title{Finding Habitable Exoplanets with Binary Relative Astrometry: Planet Detection and Characterization with the Microarcsecond Astrometric Retrieval Algorithm (MARA)}

\author[0009-0004-6306-9715,sname=Roberson,gname=William]{William Roberson}
\affiliation{Department of Astronomy, New Mexico State University, P.O. Box 30001, MSC 4500, Las Cruces, NM 88003, USA}
\email{wcr@nmsu.edu}

\author[0000-0001-6975-9056,sname=Nielsen,gname=Eric L.]{Eric L. Nielsen}
\affiliation{Department of Astronomy, New Mexico State University, P.O. Box 30001, MSC 4500, Las Cruces, NM 88003, USA}
\email{nielsen@nmsu.edu}

\author[0000-0002-8035-4778,sname=Christiansen,gname=Jessie L.]{Jessie L. Christiansen}
\affiliation{NASA Exoplanet Science Institute, IPAC, MS 100-22, Caltech, 1200 E. California Blvd, Pasadena, CA 91125}
\email{jessiec@caltech.edu}

\author[0000-0002-1871-6264,sname=Vasisht,gname=Gautam]{Gautam Vasisht}
\affiliation{Jet Propulsion Laboratory, California Institute of Technology, 4800 Oak Grove Dr., Pasadena, CA 91109, USA}
\email{gautam.vasisht@jpl.nasa.gov}

\author[0000-0002-9408-8925,sname=Bendek,gname=Eduardo]{Eduardo Bendek}
\affiliation{NASA Ames Research Center, Moffett Field, CA 94035, USA}
\email{eduardo.bendek@nasa.gov}

\author[0000-0003-2082-5176,sname=Davis,gname=Alex]{Alex Davis}
\affiliation{Jet Propulsion Laboratory, California Institute of Technology, 4800 Oak Grove Dr., Pasadena, CA 91109, USA}\email{alex.davis@jpl.nasa.gov}

\author[0000-0003-2008-1488,sname=Mamajek,gname=Eric E.]{Eric E. Mamajek}
\affiliation{Jet Propulsion Laboratory, California Institute of Technology, 4800 Oak Grove Dr., Pasadena, CA 91109, USA}\email{eric.mamajek@jpl.nasa.gov}

\author[0000-0002-2361-5812,sname=Clark,gname=Catherine A.]{Catherine A. Clark}
\affiliation{NASA Exoplanet Science Institute, IPAC, MS 100-22, Caltech, 1200 E. California Blvd, Pasadena, CA 91125}
\email{clarkc@ipac.caltech.edu}

\author[0000-0001-5253-1338,sname=Kratter,gname=Kaitlin M.]{Kaitlin M. Kratter}
\affiliation{Department of Astronomy and Steward Observatory, University of Arizona, Tucson, AZ 85721, USA}
\email{kkratter@arizona.edu}

\author[0000-0002-7733-4522,sname=Becker,gname=Juliette]{Juliette Becker}
\affiliation{Department of Astronomy, University of Wisconsin-Madison, 475 N Charter St, Madison WI 53706, USA}
\email{juliette.becker@wisc.edu}

\author[0000-0003-1227-3084,sname=Meyer,gname=Michael R.]{Michael R. Meyer}
\affiliation{University of Michigan, Department of Astronomy, 1085 South University, Ann Arbor, MI 48109, USA}
\email{mrmeyer@umich.edu}

\begin{abstract}
Binary relative astrometry is a technique to search for rocky planets in the habitable zone of nearby binary stars using 1D relative astrometry at the microarcsecond level. This unprecedented precision would allow a custom-designed space telescope to directly measure the occurrence rate of these planets. The success of such a mission depends on our ability to recover and characterize planets from the unique format of extreme precision binary relative astrometry data. We present MARA, the Microarcsecond Astrometric Retrieval Algorithm, specifically designed for these data. We describe the design and format of the MARA pipeline, and demonstrate its accuracy and performance with a series of validation tests on simulated data, using the SHERA SMEx mission concept as an example. Our injection/recovery tests show that with these data, MARA is able to detect and characterize rocky planets in the habitable zone of $\alpha$ Cen A, down to a coplanar mass of $\sim$1 M$_\oplus$ in 1 year orbits. Expanding to a range of input planet masses and periods for the same example mission, we find that the results from these injection/recovery tests generally agree with the analytic predictions of binary relative astrometry sensitivity. We use MARA to map out the expected completeness as a function of planet mass and period, which in this case reaches down to $\sim$0.5 M$_\oplus$ at 3 year orbits around $\alpha$ Cen A. These depth-of-search calculations will be a vital ingredient in demographics calculations from the final data from a binary relative astrometry mission.

\end{abstract} 

\section{Introduction}\label{sec:intro}
The search for exoplanets has been extremely fruitful, with over $6000$ exoplanets found to date (e.g. \citealt{exoplanet_archive}). However, despite the large number of exoplanets, Earth-analog exoplanets (rocky planets in the habitable zones of Sun-like stars) are out of reach for most detection methods, even using state-of-the-art facilities \citep{christiansen:2026}. Additionally, while about half of Sun-like stars are in multiple systems \citep{offner:2023}, less than $10$\% of planets on the NASA Exoplanet Archive are in multiple systems. There is currently not enough data to directly measure the occurrence rate of Earth-analogs, nor determine how this occurrence rate is affected by binarity. Both of these science questions can be addressed with binary relative astrometry, a detection method where S-type planets can be detected by measuring perturbations in the separation between two stars in a binary. Multiple current mission concepts aim to use this technique to search for Earth-analogs in binary systems.

A unique capability of these binary relative astrometry missions is to find rocky planets in the habitable zone of nearby binary stars, and to determine whether the occurrence rate of  planets in binary systems is suppressed relative to the occurrence rate around single stars \citep{christiansen:2026}.  Previous work suggests the occurrence rate of closer-in mini-Neptunes and super-Earths are suppressed around binaries (e.g. \citealt{sullivan:2026}), and binary relative astrometry missions will determine if this suppression continues out to $1.5$ $AU$.

One of these missions is the Searching for Habitable Exoplanets with Relative Astrometry (SHERA) SMEx Mission Concept. The SHERA mission concept will be a 22 cm space telescope that uses a modified diffractive pupil in order to measure precision relative astrometry of 14 target stars in nearby binary systems \citep{christiansen:2026}. The diffractive pupil, which breaks up the PSF into multiple features and allows for wavefront sensing in the image plane, results in a much more accurate measurement of the relative astrometry of two stars in a binary (e.g. \citealt{bendek:2013}, \citealt{langford:2024}). SHERA will be able to measure the separation of nearby binaries to $\sim$1 microarcsecond (\uas) precision, which is sufficient to detect the astrometric deflection caused by rocky planets with masses as low as $1$ $M_\oplus$ around the closest binaries. SHERA will monitor these targets for 3 years, achieving sensitivity to rocky planets across the entire habitable zones of these stars. SHERA will also perform valuable preparatory work for HWO, as 13 of the $14$ SHERA targets are Tier 1 targets for HWO, and the last is a Tier $2$ target \citep{mamajek:2024,harada:2024}.

Another binary relative astrometry mission concept is the Telescope for Orbital Locus Interferometric Monitoring of our Astrometric Neighborhood (TOLIMAN) concept \citep{tuthill:2026}. TOLIMAN will also use a modified diffractive pupil, but on a 12.5 cm spacecraft. While SHERA will survey multiple binaries, TOLIMAN will concentrate on the $\alpha$ Cen AB system, seeking to detect rocky planets orbiting either star over a 3 year mission with $\sim$1 \uas\ precision. Beyond these two missions, the diffractive pupil concept presents an opportunity to reach much smaller planets in the habitable zones of nearby binaries compared to any other current technique.

An important component of the success of a binary relative astrometry mission is the planet detection and characterization algorithm. Due to the high precision and the unique form of the 1D astrometric data additional care is needed for orbit modeling. Here, we demonstrate that period and mass can be recovered to sufficient precision, that multiple planetary signals can be disentangled reliably, and that binary relative astrometry data will be sensitive to rocky planets in the habitable zones of nearby stars. To accomplish this, we have developed and tested planet detection and characterization algorithms: the Microarcsecond Astrometric Retrieval Algorithm (MARA), which we present here.

In Section \ref{sec:architecture}, we describe the format of binary relative astrometry data and MARA's architecture. In Section \ref{sec:validation}, we discuss the validation tests we have run on MARA, with example simulated datasets based on the SHERA mission concept. In Section \ref{sec:diss}, we present our sensitivity analysis from MARA, and discuss the example of SHERA observations of $\alpha$ Cen A. In Section \ref{sec:conclusion} we discuss future work and summarize our results.

\section{MARA Architecture}\label{sec:architecture}
\begin{figure*}
    \centering
    \includegraphics[width=0.99\textwidth]{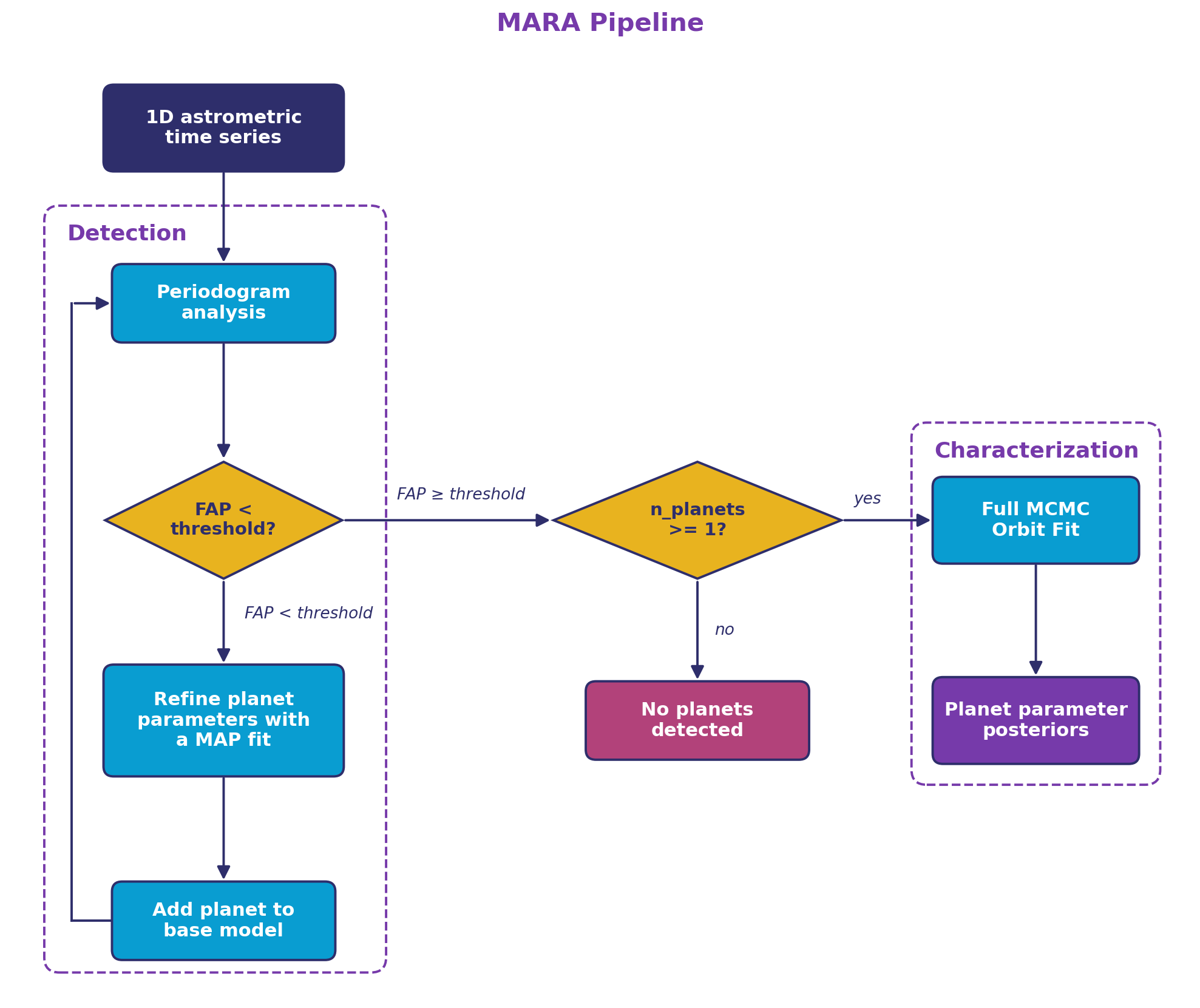}
    \caption{The overall architecture of the MARA detection pipeline for binary relative astrometry data, which separates detection and characterization. An initial periodogram search determines if there is a planet above the threshold level of $10^{-3}$ False Alarm Probability (FAP), and a maximum $a$ $posteriori$ (MAP) fit allows that planet to be removed from the dataset. The process is repeated until no signals remain above the set threshold. Once one (or more) planets are identified from this planet detection, an MCMC fit returns the planet parameter posteriors.}
    \label{fig:flowchart}
\end{figure*}

The planet detection and characterization pipeline for any binary relative astrometry mission is crucial for both analyzing in-flight data and planning the survey. We follow a similar method to that used by the California Legacy Survey (CLS) in the \texttt{RVsearch} module \citep{fulton:2018}, dividing detection and characterization into two separate steps. This allows us to run the more computationally-intensive characterization method on a small subset of simulated datasets to verify that the detection and characterization methods are consistent. We are then able to run the much faster planet detection step on tens of thousands of simulated datasets to measure sensitivity. The flow of the pipeline is shown in Figure \ref{fig:flowchart}. Following \citet{fulton:2018}, we perform a periodogram analysis, followed by a maximum $a$ $posteriori$ fit, for each planet detected in the data. Once planets are detected, a Markov Chain Monte Carlo fit is performed to recover posteriors on planet parameters.

\subsection{Format of Binary Relative Astrometry Data}
A planet orbiting a nearby star can be detected from the astrometric reflex motion of the star around the planet/star center of mass (e.g. \citealt{perryman:2016,lammers:2026}). For two stars in a binary, one of which has an orbiting S-type planet, the planet is detectable from the changes in the relative astrometry between the two stars. Binary relative astrometry missions will measure the separation between two stars in nearby binary systems, in the plane of the sky, at the \uas\ 
level. As these missions will measure separation to much higher precision than position angle, this will be essentially a one-dimensional measurement (though there will be some orbital motion of the binary over a mutli-year mission). This makes binary relative astrometry data somewhat analogous to detecting planets from stellar radial velocities (RVs), since orbital parameters must be extracted from a 1D projection of 3D motion.

\begin{figure*}
    \centering
    \includegraphics[width=0.99\textwidth]{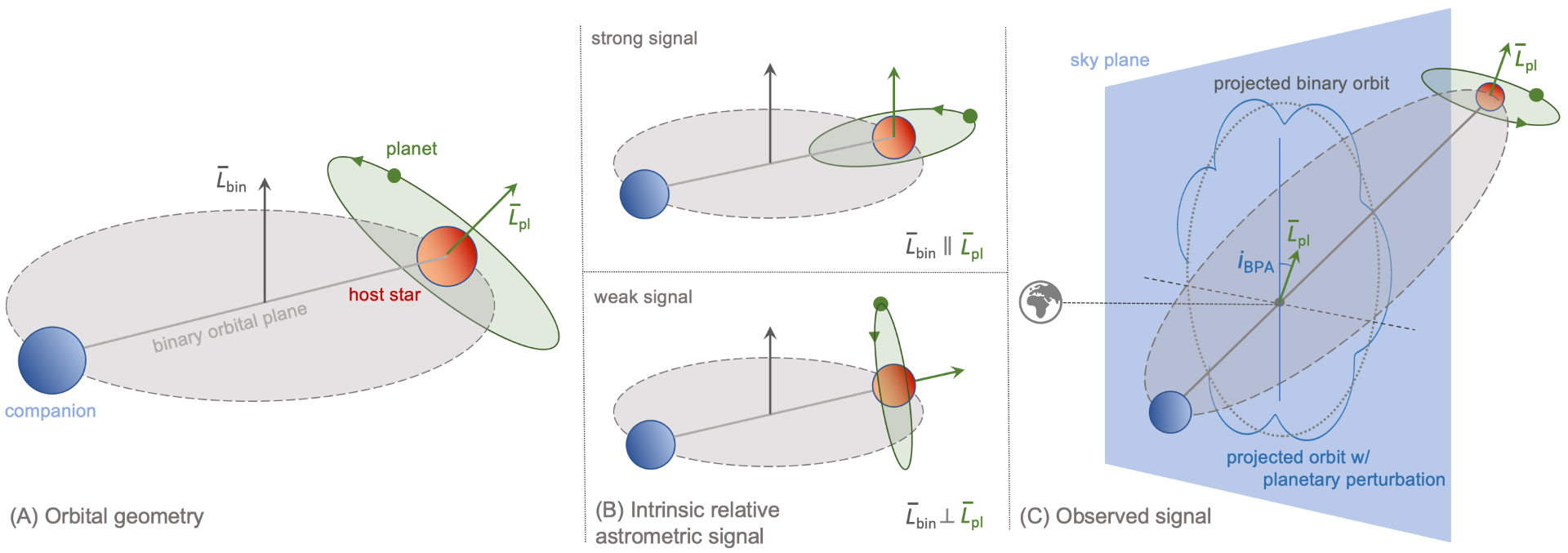}
    \caption{Since a binary relative astrometry mission will make one-dimensional measurements of the deflection caused by the planet on the separation of two stars in a binary, the relative orientation of the binary stars' angular momentum vector and that of the planet's orbit (panel A) sets the strength of the signal. A set of coplanar orbits (top of Panel B) will produce one of the strongest astrometric signals. At the other extreme, an orientation where very little of the stellar reflex motion is along the line connecting the two binary stars will produce a much weaker signal (bottom of Panel B). Since only the projected motion is observed (Panel C), the full 3D orientation of the orbit cannot be recovered from binary relative astrometry data alone, similar to the case with RV-detected planets. Additional observations of the same targets, such as with extreme precision radial velocity, would break this degeneracy.}
    \label{fig:orbit_orientation}
\end{figure*}

Extending this analogy, the full 3D orbital parameters cannot be determined from these 1D data alone, in much the same way inclination angle and position angle of nodes are not recovered with RV measurements. While the relevant angle for RVs is the extent to which the star's motion is projected onto the line of sight, for binary relative astrometry the key is what fraction of the star's reflex motion is projected onto the line along the plane of the sky connecting the two stars in the binary. As shown in Figure~\ref{fig:orbit_orientation}, the strength of the signal will depend on the ($a$ $priori$ unknown) orientation between the angular momentum vectors of the binary stars and the planet's orbit. A planet in an edge-on orbit (as seen from Earth) whose orbital angular momentum vector points along the line connecting the two binary stars in the plane of the sky will induce the smallest astrometric signal. The strongest signal, then, will come from a planet whose orbital angular momentum vector is perpendicular to this line. Generally, a planet coplanar with the orbit of the binary will induce a relatively strong signal.

Binary relative astrometry data will at its essence be a time series of separations of the two stars in a binary. These data will encode both the properties of any orbiting planets, but also the binary orbital motion, the system proper motion and radial velocity, and spacecraft motions within the Solar System itself \citep{christiansen:2026}. We consider three levels of binary relative astrometry data in our analysis below:

\begin{itemize}
    \item Level 1: the separations as directly measured by the spacecraft, including planet signal, binary orbit, target binary space motion, parallax, spacecraft motions, and relativistic effects.
    \item Level 2: Parallax, spacecraft motions, and relativistic effects are removed, as though the data were taken from the Solar System barycenter.
    \item Level 3: Binary orbit and space motion are also removed, leaving behind only the reflex motion caused by the planet.
\end{itemize}

\noindent While ultimately science from binary relative astrometry missions will involve working with Level 1 data, for prototyping, testing, and pipeline development purposes we concentrate on Level 2 and Level 3 data. The key difference between Level 2 and Level 1 data will be fitting for spacecraft motions, system parallax, the differential parallax between the stars, and relativistic effects. Differential parallax is the parallax difference between the two stars in the binary. For example, $\alpha$ Cen A will be about 2.5 AU closer to Earth in 2031 compared to $\alpha$ Cen B. They will be at the same distance in 2031.9, and B will be about about 5.5 AU closer than A in 2034 \citep{akeson:2021}. This corresponds to a differential parallax signal of $\sim$7, $\sim$0, and $\sim$15 \uas\, respectively. This differential parallax will have the largest impact, assuming the spacecraft position and velocity are known precisely (for example, SHERA's position will be known to $10$ $m$ and its velocity to $5$ $cm /s$), because the system parallax will not change the apparent separation of the stars. Ultimately, a binary relative astrometry mission could measure differential parallax to $\sim 1$ \uas, but this signal would have to be fit alongside the planet signal. We therefore expect the main impact from switching from Level 2 to Level 1 data will be a decreased sensitivity to planets with similar period ($\sim$1 yr) and phase to the differential parallax signal, but further study will be required, especially to understand to what precision a constant amplitude planet signal can be separated by a changing amplitude differential parallax signal. While we defer the more detailed Level 1 fits to future work, here we investigate general properties of the data, the fitting process, and constraints on planets from Level 2 and Level 3 data. As we demonstrate below, while it is more computationally expensive to fit Level 2 data, the planet posteriors we recover from fits to Level 2 and Level 3 data are very similar. Therefore we mainly use the example of simulated Level 3 data appropriate to the SHERA mission concept in our description of the pipeline below.

\begin{figure*}[t]
    \centering
    \includegraphics[width=\textwidth]{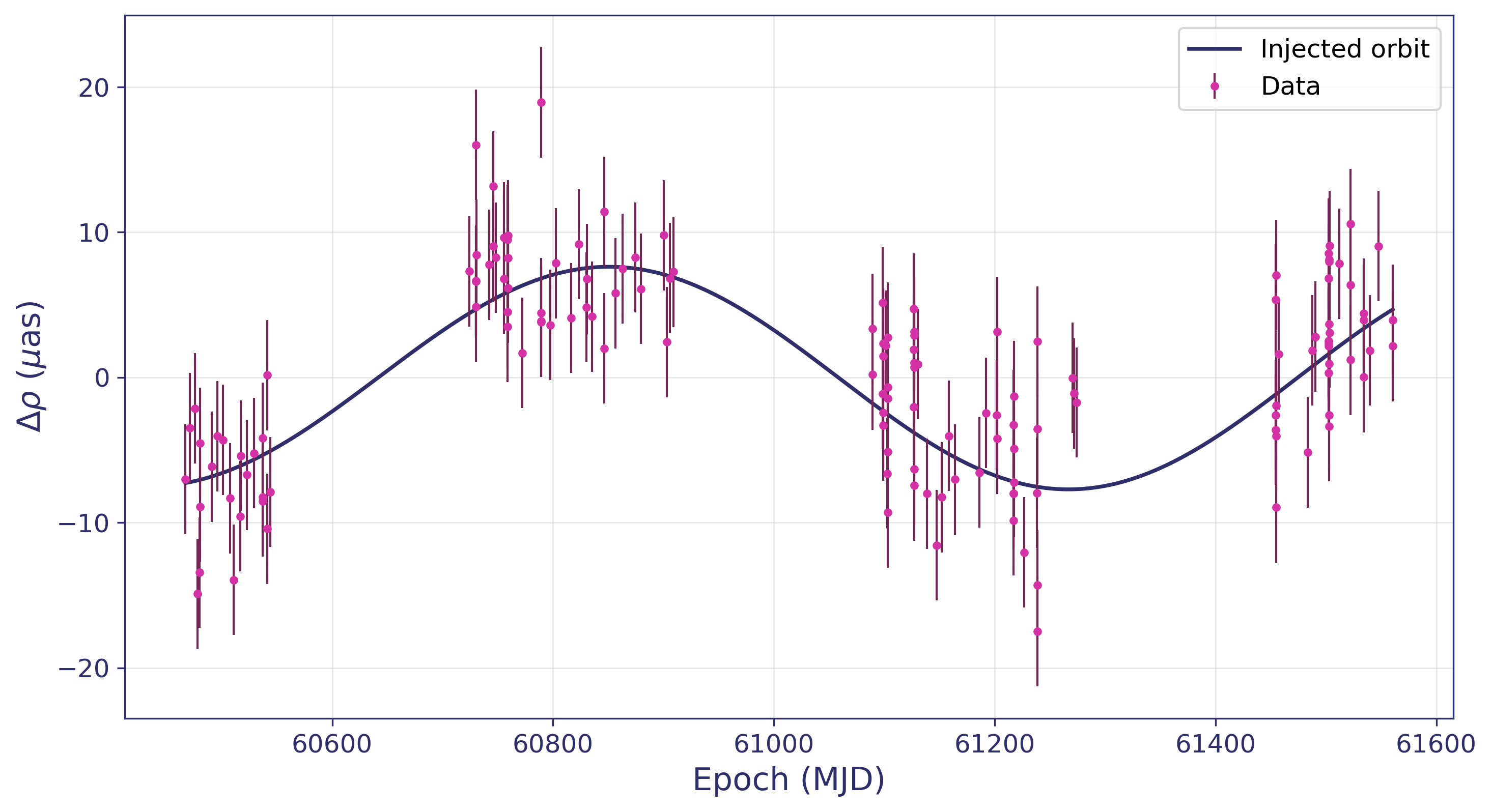}
    \caption{Simulated binary relative astrometry data, from the proposed SHERA mission, of a 2.27 year, 2.11 M$_\oplus$ planet orbiting $\alpha$ Cen A. Over 3 years, the separation between the two stars in the binary changes by $\sim10$ \uas\, 
    which is detectable with the 4 \uas\ precision measurements of SHERA. The MARA pipeline is designed to extract planet masses and orbital parameters from these data.}
    \label{fig:simulated_data_example}
\end{figure*}

An example of this simulated SHERA Level 3 data is shown in Figure~\ref{fig:simulated_data_example}, for a 2.11 M$_\oplus$ planet in a 2.27 year orbit around $\alpha$ Cen A. These data are generated using astrometric and orbital parameters of $\alpha$ Cen from \citet{akeson:2021}, and SHERA observing simulations \citep{christiansen:2026} in 30 minute observing blocks. Scatter in these datasets is simulated that includes $\sim 4$ \uas\, measurement precision as well as other sources of noise. These noise sources include surface variability in $\alpha$ Cen A and B, background star contamination, instrumental variation, and detector artefacts \citep{christiansen:2026}. The gaps in the dataset correspond to the several months each year the system is too close to the Sun, and therefore unobservable due to SHERA's 90$^\circ$ Solar Keepout restriction.

\subsection{Orbit Fitting Methodology}

As we detail below, both the planet detection step and the characterization step utilize a specialized orbit fitting method. This is due both to the format of binary relative astrometry data, 1D measurements of separation between the two stars in the binary, and the high precision. At the $\sim$1 \uas\ 
level, several effects that can normally be neglected in astrometric measurements become relevant. Most important are the finite speed of light and perspective effects (the difference between spherical coordinates and Cartesian coordinates). To model orbital parameters from binary relative astrometry data we adapt the \texttt{orbitize!} orbit fitting Python module (\citealt{orbitize1}; \citealt{orbitize2}; \citealt{ptemcee}). Our modifications both address these higher order effects and increase fitting efficiency.

We split out the orbit fitting into two parts, for Level 3 binary relative astrometry data (where only the planet signal remains) and Level 2 data (the planet signal, binary orbit, and system parallax and proper motion are present). Both types of fits require five parameters to model each planet, and eight parameters to model the orbit and masses of the binary stars, as well as the system parallax, as detailed in Table~\ref{tab:fit_params}.

\begin{table}[h]
\centering
\begin{tabular}{lll}
\hline
\textbf{Fit Parameter} & \textbf{Symbol (Unit)}  & \textbf{Data Level} \\
\hline
Binary Primary Mass                & $M_A$ ($M_\odot$)  &2,3\\
Binary Period                      & $P$ (yr)  &2,3\\
Binary Semimajor Axis              & $a_b$ ($^{\prime\prime}$)  &2,3\\
Epoch of Periastron Passage        & $t$ (MJD)  &2,3\\
Orbital Eccentricity               & $e$  &2,3\\
Inclination Angle                  & $i$ ($^{\circ}$)  &2,3\\
Argument of Periastron             & $\omega$ ($^{\circ}$)  &2,3\\
Longitude of Ascending Node        & $\Omega$ ($^{\circ}$)  &2,3\\
\hline
 Parallax& $\pi$ ($^{\prime\prime}$)&2,3\\
 Reference RA& $\alpha_0$ ($^{\circ}$)&2\\
 Reference Dec& $\delta_0$ ($^{\circ}$)&2\\
 RA Proper Motion& $\mu_\alpha^*$ (mas/yr)&2\\
 Dec Proper Motion& $\mu_\delta$ (mas/yr)&2\\
 System RV& $RV$ (km/s)&2\\
\hline
Planet $n$ Semimajor Axis          & $a_n$ (AU)&2,3\\
Planet $n$ Eccentricity            & $e_n$&2,3\\
Planet $n$ Argument of Periastron  & $\omega_n$ ($^{\circ}$)&2,3\\
Planet $n$ Phase        & $\tau_{\mathrm{PA},n}$&2,3\\
Planet $n$ Coplanar Mass& $m_{coplanar,n}$ ($M_\oplus$)&2,3\\
\hline
\end{tabular}
\caption{Orbit fitting parameters for the binary and planetary orbits. Fits to both Level 2 and Level 3 data involve 8 parameters from the binary orbit and the system parallax, as well as five parameters per planet. An additional 5 parameters for the system are required for fits to Level 2 data.
\label{tab:fit_params}}
\end{table}

Level 3 data are modeled as the amount of reflex motion by the host star due to the planet, along the line connecting the two binary stars in the plane of the sky. This simplifies to a dot product between offset of the host star from the host star/planet center of mass and the normalized vector from primary to secondary in the plane of the sky:

\begin{equation}
\Delta \rho = {\rm cos}(PA_{Ap,n} - PA_{AB})\sqrt{\alpha_n^2 + \beta_n^2} 
\end{equation}

\noindent where $\alpha_n$ and $\beta_n$ are the RA and Dec offset of the host star from the host star/planet center of mass due to planet $n$, $PA_{AB}$ is the position angle of the secondary with respect to the primary in the stellar binary, and $PA_{Ap,n}$ is the position angle of planet $n$ with respect to the primary, at the time of a given observation. Level 2 data requires 5 additional parameters: RA and Dec of the system barycenter at a reference epoch, the system proper motion in RA and Dec, and the system radial velocity. These additional parameters are needed to properly model the 3D motion of the system.

While light travel time and perspective effects can be neglected in Level 3 data, they must be included when fitting Level 2 data. The 3D motion of the system affects the separation between the two stars at the \uas\
level. With a radial velocity of about $-22.4$ \kms\ and parallax of $\sim$750 mas \citep{akeson:2021} the $\alpha$ Cen system is about 2400 light seconds closer to Earth with each passing year. Put another way, the parallax is increasing by about 13 \uas\
a year. Given the orbit of the two stars, a 2400 second timing error in the middle of the mission observation baseline translates to a $\sim$65 \uas\
separation error.

We correct for light travel time through an iterative process. We first calculate the orbital position (of both the stellar binary and the host star about the host star/planet center of mass) at the observed epoch. These 3D positions give the distance offset (in light seconds) between each star and the system barycenter at a reference epoch. The choice of reference epoch is arbitrary, and is most naturally the epoch used for the measurement of orbital properties and system space motion for a given binary system. For $\alpha$ Cen, these measurements come from \citet{akeson:2021}, who use 2019.5 as the reference epoch, which we also adopt for fits to simulated $\alpha$ Cen data. The observing times are then updated based on these time offsets, and the calculations are repeated. From the second to the third iteration, the separation changes by $\lesssim 0.0015$ \uas,
 and so we find that only two iterations are required.

At this level, perspective effects are also relevant. Generally, RA, Dec, and distance are treated as a pseudo-Cartesian set of coordinates for a single system. However, for nearby stars with large proper motion and parallaxes (e.g. the TOLIMAN and SHERA targets), their angular velocity is large enough that this assumption breaks down. In particular, while the total space motion (the amplitude of the 3D velocity vector) can still be assumed to be constant over time, the projection of this vector into RA, Dec, and distance changes over time (e.g. \citealt{lindegren:2021}). This effect is the origin of secular acceleration seen in precise RV orbit fits (e.g. \citealt{zechmeister:2009}). This is particularly relevant here since a binary relative astrometry mission measures separation along the plane of the sky, but the orientation of the plane of the sky shifts over time. During 2031, this effect will change the separation between $\alpha$ Cen A and B by $\sim120$ \uas. Another way to interpret this is that the traditional orbital elements (especially the angles $i$, $\omega$, and $\Omega$) are not constants, but rather slowly vary with time.

We correct for perspective effects by taking our orbital parameters to refer to a specific reference epoch (2019.5 for $\alpha$ Cen).  At each observing epoch the RA and Dec offset of the two stars, and the offset of the host star from the host star/planet center of mass, are calculated. The resulting RA, Dec, and distance (spherical coordinates) are then converted to $X,Y,Z$ Cartesian coordinates, as are the proper motions and parallaxes. The motion of the system and the orbit perturbations are then performed in Cartesian coordinates, before transforming back to spherical coordinates to calculate the final offset between the two stars on the plane of the sky.

The next modification to \textit{orbitize!} is to match the observable returned by the orbit model to binary relative astrometry data: one-dimensional separation. In fits to Level 2 data this is the total separation between the two stars, and in Level 3 fits it is just the separation offset caused by the planet.

The final modification is adding new basis sets that are more efficient for modeling binary relative astrometry data. Continuing the analogy to radial velocity, RV exoplanet fits are most often performed in the basis of velocity semi-amplitude, eccentricity, period, argument of periastron, and epoch of periastron passage (e.g. \citealt{fulton:2018}). While an RV orbit can instead be fit with a basis appropriate to direct imaging (semi-major axis, eccentricity, inclination angle, argument of periastron, position angle of nodes, epoch of periastron passage, and total mass), such a fit is much less efficient. For example, period is often strongly constrained by an RV dataset (especially if multiple orbits have been observed), but a fit must adjust both semi-major axis and total mass simultaneously to dial in a particular period. Similarly, inclination angle and position angle of nodes are not constrained by RV orbits, so including these parameters slows the fit. Other basis changes speed convergence for circular orbits, such as fitting in $\sqrt{e}\sin \omega$ and $\sqrt{e}\cos \omega$ rather than eccentricity and argument of periastron directly \citep{ford:2006}, or fitting in time of conjunction rather than epoch of periastron passage.

Working with simulated binary relative astrometry data, we have found the most efficient basis set to be planet semimajor axis ($a,$ in AU), eccentricity ($e$), argument of periastron ($\omega$, degrees), time of binary PA ($\tau_{PA}$), and coplanar mass (the mass the planet would have if it is in the same orbital plane as the binary, in Earth masses). Planet semi-major axis (along with mass of planet and mass of the host star) sets the semi-major axis of the reflex motion of the host star:

\begin{equation}
    a_* = \frac{m_p}{M_* + m_p} a_p
\end{equation}

\noindent which, for a particular orientation of orbit, sets the amplitude of the planet signal in the binary relative astrometry data. Eccentricity and argument of periastron, as in RV orbits, set the shape of the curve in $\Delta \rho$; for a perfectly circular orbit where the binary had negligible orbital motion over the observing window, this would be a perfect sine curve.

We choose to parameterize the phase of the orbit with $\tau_{PA}$, as this corresponds more directly to the observed phase from the data itself. Also, unlike epoch of periastron passage, $\tau_{PA}$ is well-defined even in the case of a circular orbit (similar to time of conjunction in RV orbit fits). $\tau_{PA}$ is a dimensionless quantity that is 0 when the planet's position angle (as seen from Earth) is equal to a reference PA, which we set to be the expected PA of the binary stars at the midpoint of the observations plus an additional 90$^\circ$. If the planet were coplanar with the binaries, this would be the moment when the planet is at a right angle with the line connecting the two stars in 3D space. Following the example of the $\tau$ parameter in \texttt{orbitize!} \citep{orbitize1}, the quantity is normalized by the planet's orbital period, so $\tau_{PA} = 0.75$ is when the planet is closest to the other star in the binary, and $\tau_{PA}=0.25$ is when the planet is furthest, for circular orbits. For example, in Figure~\ref{fig:simulated_data_example}, $\tau_{PA}=0$ represents when the curve crosses 0 from positive values to negative values: the planet is moving away from the other star in the binary, so the host star is moving toward the other star.

Given the 1D nature of the observations, we do not fit directly for planet mass, $m_p$, but rather coplanar mass, $m_{coplanar}$. That is, the mass the planet would have, given the amplitude of the astrometric signal, if the planet's and the binary's orbital planes were the same. Fitting in coplanar mass avoids the degeneracy between the actual mass of the planet and the relative orientation of the two orbital planes.

Returning to the RV analogy, since RV observations do not measure inclination angle, $m_p \sin i$ is reported instead of planet mass, representing the minimum planet mass. The relevant angle for binary relative astrometry is not inclination angle (the angle between the planet's orbital angular momentum vector and the line of sight), but rather an angle we designate $i_{BPA}$ (panel C of Figure~\ref{fig:orbit_orientation}). We define this angle as the angle between the angular momentum vector of the planet's orbit and the line in the plane of the sky connecting the two stars in the binary, at a chosen reference epoch. We take this reference epoch to be the midpoint of the observations. In the same way that a planet with $i=0^\circ$ will induce no RV signal on the star since all the orbital motion is perpendicular to the line of sight, a planet with $i_{BPA} = 0^\circ$ will not change the observed separation between the two stars, since all the orbital motion is perpendicular to the line connecting the stars on the plane of the sky.

There is a small difference for the TOLIMAN and SHERA targets, since orbital motion will slowly change the PA of the binary over the observing window (for example, $\alpha$ Cen AB is expected to move by $\sim$10$^\circ$ between 2031 and 2034), but it remains true that the smallest possible astrometric signal will occur for an orbit with $i_{BPA} = 0^\circ$ or $180^\circ$, whereas the maximum signal will be at $i_{BPA} = 90^\circ$. Such a maximum signal (corresponding to the minimum possible planet mass) is generally not a coplanar orbit, unless the binary star's orbit has an inclination angle of $i=0^\circ$. Nevertheless, a coplanar orbit will generally correspond to a relatively large 1D astrometric signal. Computationally, there is not a large difference between fitting in $m_p \sin i_{BPA}$ and $m_{coplanar}$. In an MCMC fit it is trivial to convert between the two, since the star's and planet's orbital parameters are known for each step of the chain. As a result, we choose to work in coplanar mass since it has the most direct physical meaning, and does not depend on observational epoch.

\subsection{Planet Detection: Periodogram and MAP Fit}

Following \citet{fulton:2018}, we begin with a periodogram method to determine if a planet is present in a binary relative astrometry dataset, before proceeding to the more computationally expensive MCMC fitting. We use a Lomb-Scargle periodogram with floating mean enabled to measure periodogram power, with a frequency sampling of $(2 \pi T_{baseline})^{-1}$
where $T_{baseline}$ is the observational baseline, as described in \citealt{cls}. A threshold false alarm probability (FAP) of $10^{-3}$ is adopted, based on comparison with MCMC results described below. If the highest peak in the periodogram exceeds a FAP of $10^{-3}$, a planet with the corresponding period is considered to be detected.

\begin{figure*}[t]
    \centering
    \includegraphics[width=\textwidth]{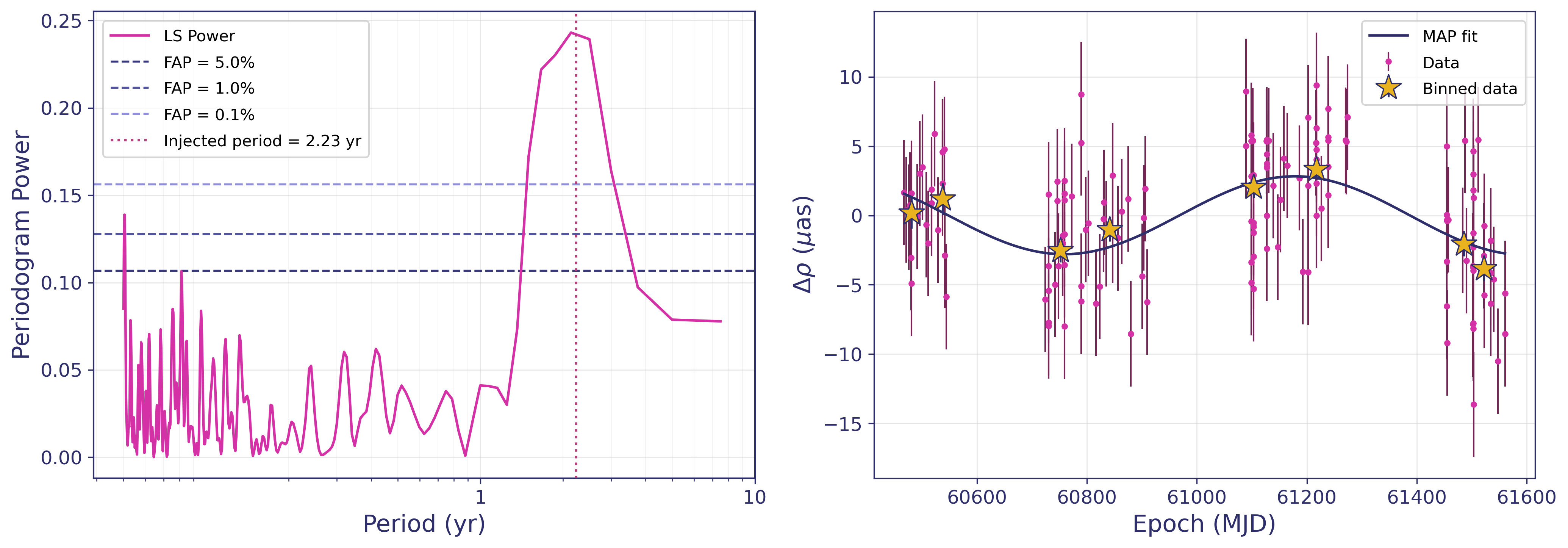}
    \caption{Periodogram analysis of simulated Level 3 SHERA data of a planet orbiting $\alpha$ Cen A (left) and the Maximum $a$ $posteriori$ (MAP) fit to the data (right). The highest peak of the periodogram corresponds to the period of the injected planet, and exceeds the threshold false alarm probability of $10^{-3}$; the FAP of this signal is $10^{-6}$. The MAP fit, while not capturing the multidimensional posterior, provides a reasonable starting estimate of the planet parameters responsible for the observed data.}
    \label{fig:periodogram_and_map_fit}
\end{figure*}

Next, we refine the period of the signal by performing a $\chi^2$ minimization over a subgrid of periods near the periodogram peak. At each of nine periods spread from 1\% below to 1\% above the periodogram peak, we vary $e$, $\omega$, and $\tau_{PA}$ to find the minimum $\chi^2$ for each period. We take the refined period as the subgrid point with the lowest $\chi^2$.
We then perform a Maximum $a$ $posteriori$ (MAP) fit to the data, allowing the parameters for all planets to vary. This fit is performed using the orbit model code described above and a $\chi^2$ minimizer. The results for Level 3 simulated data, from SHERA, for a 0.813 M$_\oplus$ planet in a 2.226 year orbit around $\alpha$ Cen A, are shown in Figure \ref{fig:periodogram_and_map_fit}. There is a clear maximum in the periodogram at the injected period, and the MAP fit comes very close to the injected signal, recovering a 0.774 M$_\oplus$ planet at 2.292 years.

After a planet has been detected, we search for additional planets in the system. The planet orbit found by the MAP fit is subtracted from the data, and the above process is repeated, analogous to \texttt{RVsearch}. When the MAP fit is carried out for the cases with 2 or more planets, all planet parameters (including those for planets identified in previous steps) are allowed to vary. This continues until no signal in the periodogram is found above the FAP threshold of $10^{-3}$. After the detection method is complete, the planets and their estimated parameters are returned. For a single data set this process takes $<$1 second, making it rapid enough to use for the injection/recovery tests (with thousands of injected planets per star) that will trace out the sensitivity of a binary relative astrometry for a given target star.

\subsection{Planet Characterization: MCMC}

Following the detection and initial characterization of the planet, we solve for the posterior on the full set of orbital parameters with a Markov Chain Monte Carlo (MCMC) method. This MCMC is carried out within \texttt{orbitize!}, using the \texttt{ptemcee} module \citep{ptemcee}. For fits to both Level 2 and Level 3 data, 280 chains at 30 temperatures are run for 10,000 steps each. Convergence is assessed by comparing posteriors separated by at least 1,000 autocorrelation times and ensuring they are consistent, and unconverged chains are run for an additional 10,000 steps until convergence is reached, with starting positions drawn from posterior samples with high likelihood from the previous run.

Priors on the planet parameters are uniform in argument of periastron, time of binary PA, semi-major axis, and planet mass. Eccentricity follows the Rayleigh distribution of \citet{stevenson:2025}. The 8 parameters for the binary orbit as well as the system parallax have Gaussian priors centered on the existing measurement for each binary system, with the Gaussian $\sigma$ corresponding to the reported error bars. In the case of the $\alpha$ Cen system, these come from the detailed analysis by \citet{akeson:2021}.

\begin{figure*}[t]
    \centering
    \includegraphics[width=\textwidth]{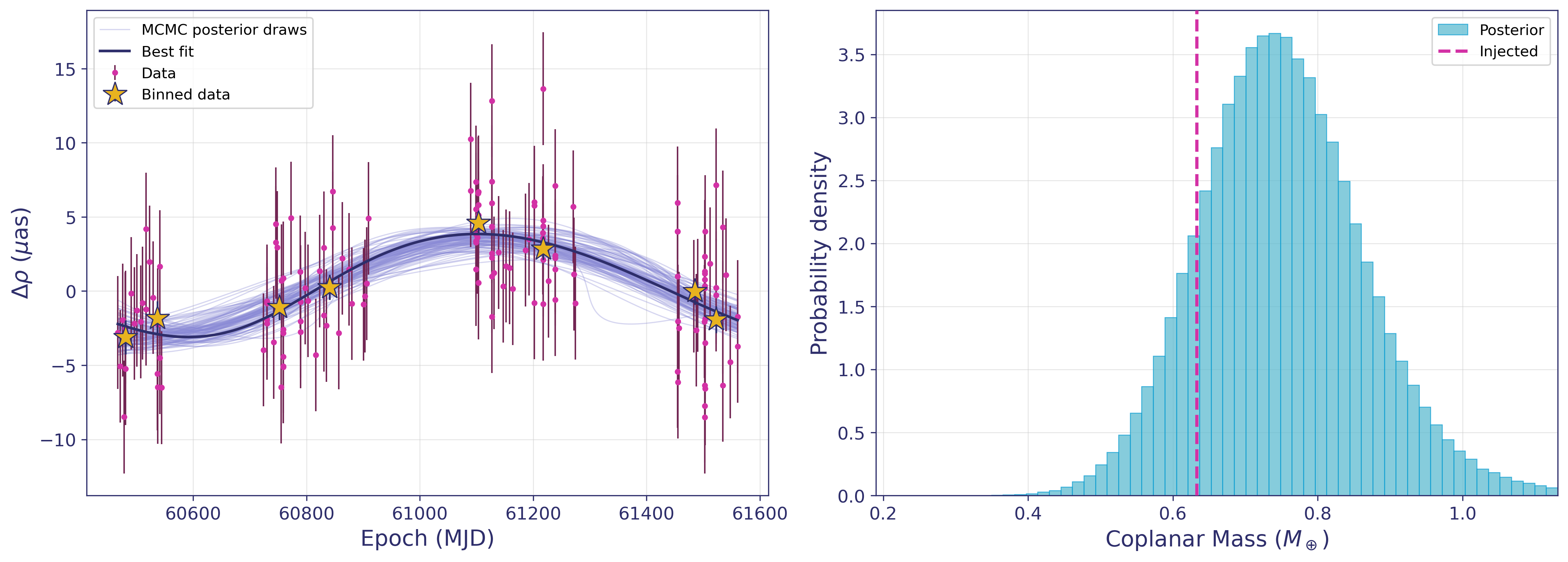}
    \caption{(left) Simulated Level 3 SHERA data for a planet orbiting $\alpha$ Cen A (red points with error bars) and orbits drawn from the posterior of an MCMC fit (blue lines), with the minimum $\chi^2$ orbit from the fit in black. There is a good match between the posterior orbits and the data, with the largest deviation generally taking place in the regions without data, which represent the Solar Keepout regions when the system is unobservable. (right) The posterior on coplanar mass, showing reasonable agreement with the input planet mass, the dashed red line. In this case, the recovered planet has a coplanar mass of $m_{coplanar} = 0.75^{+0.12}_{-0.11} M_\oplus$ and period of $P = 3.3 ^{+0.7}_{-0.4} yrs$, close to the injected value of $m_{coplanar} = 0.63$ $M_\oplus$ and $P = 2.75$ $yrs$.}
    \label{fig:one_planet_fit_and_posterior}
\end{figure*}

Figure~\ref{fig:one_planet_fit_and_posterior} shows an example MCMC fit to simulated Level 3 SHERA data, with an injected planet mass of 0.633 M$_\oplus$ and period of 2.749 years. The MCMC process results in a posterior that is a good match to the data and the input orbit. The posterior on coplanar mass ($m_{coplanar} = 0.747^{+0.115}_{-0.105}$ M$_\oplus$) is close to the input planet mass, indicating a good recovery of the planet. In a fit to Level 2 data, five additional parameters of RA and Dec proper motion, RV, and position in RA and Dec are also assumed to have Gaussian priors based on measurements, either from \citet{akeson:2021} for $\alpha$ Cen, or Gaia DR3 \citep{gaiadr3} for other systems. Uncertainty in these parameters will result in linear changes in binary separation, which is straightforward to disentangle from signals with periods less than the mission baseline.

The full posterior includes a marginalized posterior on coplanar mass (the mass the planet would have if it orbits in the same plane as the binary), as shown in Figure~\ref{fig:one_planet_fit_and_posterior}. This posterior can be transformed into a minimum mass posterior ($m_p \sin i_{BPA}$, similar to an $m_p \sin i$ posterior in any RV fit). Each step of the MCMC contains an inclination angle and position angle of nodes for the binary, allowing a direct transformation from coplanar mass to $m \sin i_{BPA}$. In the case of $\alpha$ Cen with simulated SHERA observations, these two posteriors are very similar: the posterior in Figure~\ref{fig:one_planet_fit_and_posterior} of $m_{coplanar} = 0.747^{+0.115}_{-0.105} M_\oplus$ becomes $m \sin i_{BPA}  = 0.744^{+0.115}_{-0.105} M_\oplus$.

A posterior on planet mass itself can also be generated, by assuming a uniform distribution in position angle of nodes and a $\sin i$ distribution in inclination angle. These assumptions are based on geometry if there is no correlation between the planet's orbital plane and the orbital plane of the binary. As expected, in this case the posterior is much less constrained (as would a posterior on planet mass from an RV orbit alone, as the uncertainty on viewing angle dominates). Since coplanar mass is near the minimum mass, the mass posterior gains a long tail at much larger masses. For this same orbit, the mass posterior is $m_p = 0.90^{+0.52}_{-0.18} M_\oplus$.

Finally, the nature of binary relative astrometry data creates a degeneracy in which star hosts any planet detected in 1D relative astrometry. By default, we fit planets assuming they orbit the primary star (the more massive star) in each binary. This degeneracy, however, allows a straightforward transformation of planet orbital parameters to swap the planet host from A to B, without requiring additional fitting. In fact, posteriors on all parameters remain the same, except there is a uniform offset of 0.5 in $\tau_{PA}$ (since the point between the two stars is offset by $180^\circ$, depending on which star the planet orbits), and planet semi-major axis and planet mass must be transformed, following:

\begin{equation}
    a_{p,B} = \left ( \frac{M_B}{M_A}  \right )^\frac{1}{3} a_{p,A}
    \end{equation}

\noindent and 
    
\begin{equation}
m_{coplanar,B} = \left (\frac{M_B}{M_A} \right )^{\frac{2}{3}} m_{coplanar,A} 
\end{equation}

\noindent This holds for multi-planet systems as well, so all planets can be assumed to orbit the primary star in the fitting process, with posteriors generated for the possibility they orbit the other star after the fact. For the example of the SHERA mission, this mass degeneracy is included as a term in the mission planet mass uncertainty requirement of 33\% \citep{christiansen:2026}.

\section{MARA Validation}\label{sec:validation}

\subsection{Simulated Data}
We test and validate the MARA pipeline by generating a series of simulated datasets. These datasets are generated with realistic noise  added and observing parameters appropriate to the SHERA mission concept. We use the SHERA scheduling simulator \citep{christiansen:2026} to determine the observing cadence for the generated data. This simulator includes observing gaps due to the Solar Keepout, and is designed to observe all SHERA targets regularly over the mission lifetime. We assume individual measurement errors of $3.8$ \uas, corresponding to an integrated exposure time of 30 minutes per observation.

Simulated planets are generated over a mass range from $0.05$ $M_{\oplus}$ to $3$ $M_{\oplus}$, and our simulated periods range from $0.2$ years to $6.5$ years. At this stage of validation, all planets are generated on circular orbits that are coplanar with the binary, with all planets orbiting $\alpha$ Cen A. We create simulated datasets of both Level 2 and Level 3 data to compare the fits to both. We also consider the effects of multiple planets, so that some datasets contain only one planet, while others have two planets.

\subsection{Accuracy of Recovered Parameters}
We begin our validation tests with simulated SHERA datasets containing only one planet. We analyze $10,000$ single-planet datasets with the MARA planet detection algorithm, each on circular orbits around $\alpha$ Centauri A, randomly sampled across the planet mass and period range given in the previous section. We run each through our detection pipeline, and compare the recovered planet multiplicity, masses, and periods to the injected planets. We consider a planet to be recovered if the FAP of its peak is below our threshold of $10^{-3}$.

\begin{figure*}[t]
    \centering
    \includegraphics[width=\textwidth]{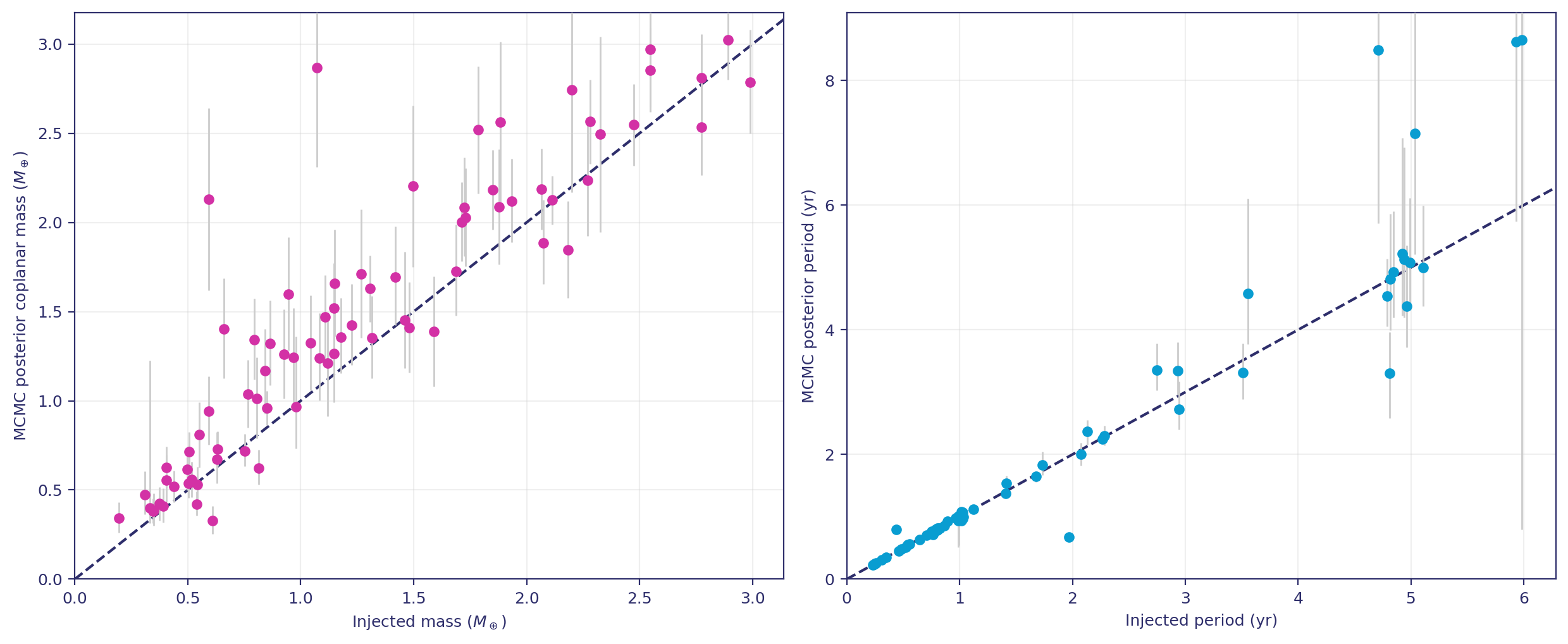}
    \caption{A comparison of the recovered coplanar mass and orbital period to the injected values for a series of 111 MCMC orbit fits with MARA to simulated 1-planet data. Overall, most MCMC posteriors overlap with the injected value at either the 1 or 2$\sigma$ confidence level, as expected. There are 5 significant outliers, corresponding to simulated datasets with particularly low SNR signals, and unfortunate time cadence gaps, which we discuss more in the text.}
    \label{fig:injected_vs_recovered_mass_period}
\end{figure*}

For a subset of 111 recovered planets we carry out more computationally expensive MCMC orbit fits as well. These datasets are chosen to straddle the boundary above and below FAP of $10^{-3}$. Comparing our MCMC posteriors on planet coplanar mass ($m_{coplanar}$) and orbital period to the injected values in the simulated datasets in Figure~\ref{fig:injected_vs_recovered_mass_period} we find generally good agreement. The injected values are within the 1$\sigma$ confidence interval of the posterior for 55\% of cases, and within 2$\sigma$ for 89\%. Typical period uncertainties (from the MCMC posterior) are $\sim$2\% for injected periods below 2 years, with uncertainties growing as the orbital period approaches and exceeds the 3-year baseline of the observations, as expected.

Of the 111 planets, 5 are significant outliers (compared to the injected value) in either mass or period. The two outliers in period result from a combination of orbital phase and period, and the timing of the Solar Keepout, resulting in an incorrect period determination, and are discussed more in Section~\ref{sec:keepout}. The three remaining outliers, with mass posteriors 3-4$\sigma$ from their injected values, all correspond to low amplitude astrometric signals, with injected masses of 0.6, 0.6, and 1.1 Earth masses at periods of 0.26, 5.0, and 0.23 years, respectively. The MCMC posteriors for these three planets correspond to mass detections between $\sim$4-5$\sigma$, suggesting additional care needs to be taken interpreting fits to planets near the detection limit, especially for very long or very short orbital periods.

There is no significant bias in the recovered period compared to the injected periods for these 111 planets. There is a small mass bias, with a linear fit to the median masses from the MCMC posterior having a slope of about 1, but lying $\sim$0.25 M$_\oplus$ above the 1-to-1 line in Figure~\ref{fig:injected_vs_recovered_mass_period}. Many of the more discrepant values correspond to lower SNR detections, while higher SNR detections mainly fall along the 1-to-1 line.

\subsection{Periodogram and MCMC Result Comparison}

While MCMC calculations of posteriors are more robust, a full MCMC fit takes about 1 day for Level 3 data, and about 2 days for Level 2 data. Periodogram analysis and MAP fits are orders of magnitude faster, taking seconds per dataset. For the 111 simulated datasets described in the previous section we compare the results from the two methods. In particular, we focus on the confidence to which planets are detected, a key consideration for any planet-finding mission. For example, a science goal of SHERA is to detect planets at a FAP $<$ $10^{-3}$. We begin by calculating the detection significance from the MCMC posterior on coplanar mass, taking a 4$\sigma$ mass measurement to correspond to a planet detection. Ideally, we would take the probability that the measured mass is less than zero, and convert that to a $\sigma$ threshold; e.g. a 5\% chance of a negative mass would correspond to a 2$\sigma$ detection. Since our mass prior prevents the MCMC from exploring negative masses, we instead compute the mass SNR as:

\begin{equation}
    SNR = \frac{m_{50}}{m_{50} - m_{16}}
\end{equation}

\noindent where $m_{50}$ and $m_{16}$ are the 50th percentile and 16th percentile confidence intervals, respectively. This is especially useful for skewed mass posteriors, with long tails toward larger masses.

\begin{figure*}[t]
    \centering
    \includegraphics[width=\textwidth]{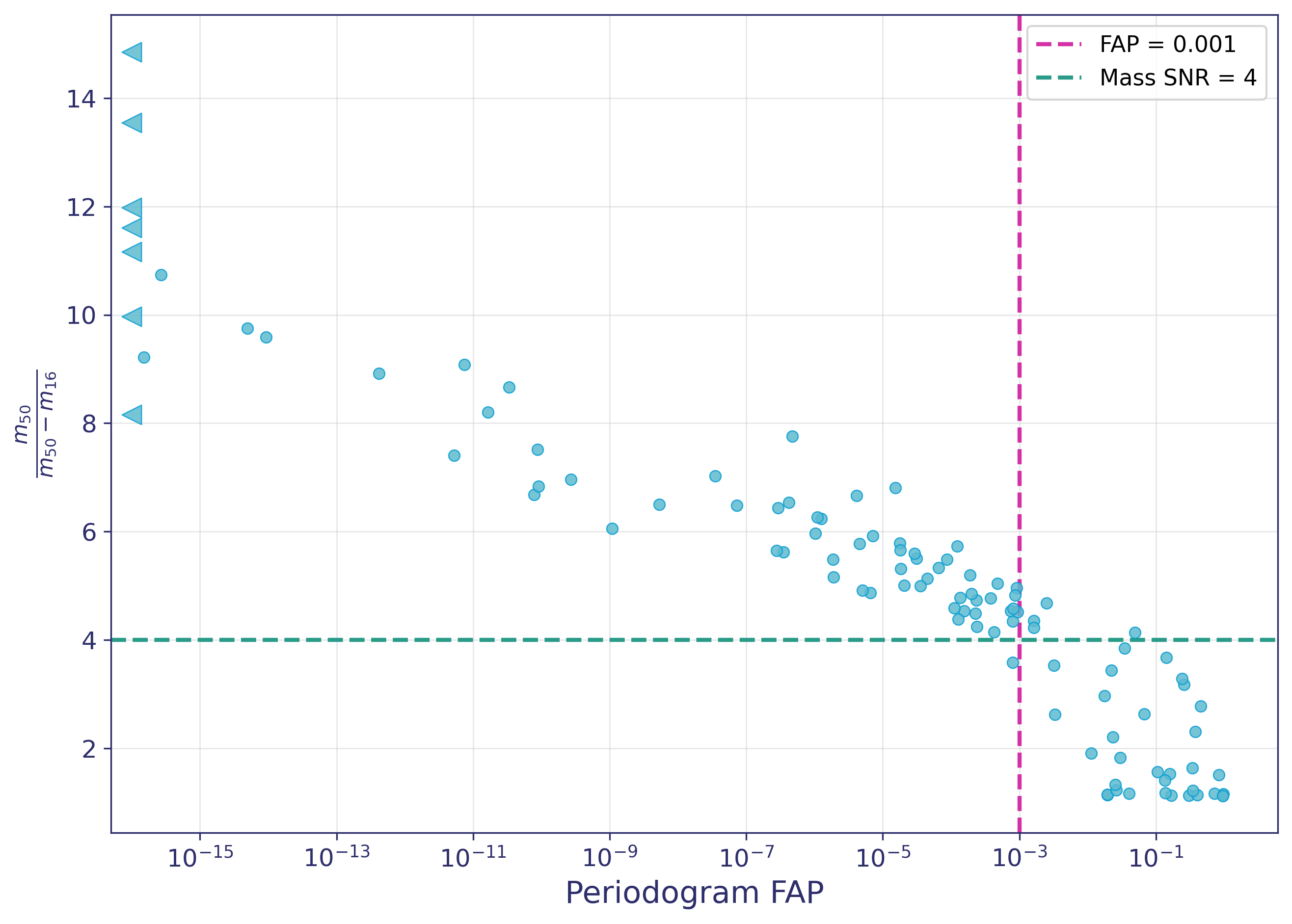}
    \caption{We find generally good agreement between the two measurements of detection significance. False Alarm Probability (FAP) from the periodogram analysis generally tracks SNR from the MCMC mass posterior, with some statistical scatter in the relationship. We adopt a threshold of $10^{-3}$ FAP which corresponds to a mass SNR of 4$\sigma$.}
    \label{fig:mass_snr_vs_fap}
\end{figure*}

Since mass posteriors can only be calculated from MCMC fits, we aim to find a proxy for a 4$\sigma$ mass detection in the computationally more efficient periodogram analysis. We compare the SNR of the detection for the 111 datasets to the initial periodogram FAP in Figure~\ref{fig:mass_snr_vs_fap}. We find a general trend, with lower False Alarm Probability corresponding to higher mass SNR, as expected. From this trend, an SNR in mass of 4 is generally reached when FAP $\le 10^{-3}$. As there is some statistical noise in this relationship, there is not an exact 1-to-1 correspondence between the two metrics, though generally the 5 cases where one threshold is crossed and the other is not are very close to at least one of the boundaries. As a result, we take a FAP of $10^{-3}$ from the periodogram analysis to correspond to a 4$\sigma$ detection of a planet when computing binary relative astrometry sensitivity to planets, while allowing for the possibility for a small amount of variation when an MCMC fit is performed on the same dataset.

The more computationally efficient periodogram analysis is especially useful for quickly identifying non-detections: when a planet is either not present in the data or significantly below the detection threshold. When we attempt to fit a dataset with no planet signal with MCMC, or attempt to fit a model with N+1 planets in a dataset with only N planets, the result is typically an unconverged MCMC process, in many cases even after several days of computation. In the case of no signal there are a very large number of possible orbits consistent with the data, resulting in an extremely long time to fully explore the allowed parameter space. When too many planets are included in the model, an additional issue slowing convergence happens when more than one chain attempts to fit the same planet signal, trading mass between them. As a result, attempting to conclude that there is a non-detection from MCMC alone is both extremely computationally inefficient and lacking in statistical robustness. We instead identify non-detections (and so the total number of detected planets per system) based on the periodogram FAP threshold.

While this reliance on the periodogram to identify detections and non-detections allows us to estimate binary relative astrometry sensitivity as a function of planet parameters, we would expect to utilize MCMC more extensively during a binary relative astrometry mission itself. Computational efficiency is an issue when attempting to fit hundreds or thousands of simulated planets, either to plan mission strategy or to compute completeness for demographics studies. Once a spacecraft is actually in flight, we would expect to run full MCMC orbit fits (with increasing numbers of planets) around every binary system the telescope observes, despite the long computation times required. In this way we would still be able to detect and characterize planets for the small number of cases where an otherwise-detectable planet happens to result in a FAP slightly worse than the $10^{-3}$ threshold.

\subsection{Impact of the Solar Keepout}\label{sec:keepout}

An important aspect of a binary relative astrometry mission is the Solar Keepout, the requirement that the telescope cannot point to close to the Sun. In the case of SHERA, the Solar Keepout prevents the telescope from pointing within 90$^\circ$ of the Sun. For most binary systems in the SHERA target list this means that a star can only be observed for $\sim$6-7 months each year, with the gaps having a uniform yearly structure. Similar to RV surveys from the ground, this results in decreased sensitivity to planets with orbits near 1 year period (or harmonics). Two of the five outliers in Figure~\ref{fig:injected_vs_recovered_mass_period} are from this effect, with injected periods of 1.97 and 0.44 years and the recovered period off by a factor of two or three.

\begin{figure*}[t]
    \centering
    \includegraphics[width=\textwidth]{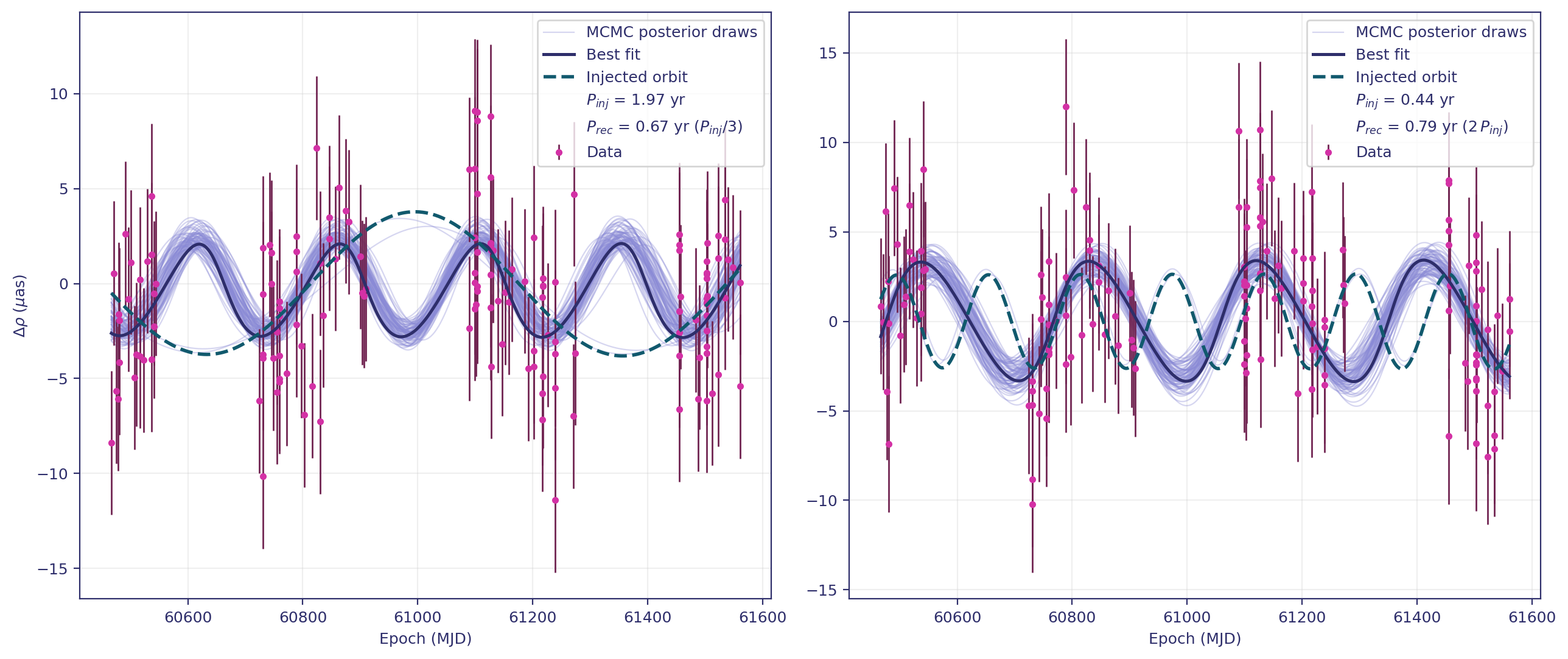}
    \caption{Two examples of fits to planets in simulated SHERA data where the median of the MCMC period posterior is a factor of 2 or 3 away from the injected period. In both cases, the period of the planet is close to a harmonic of 1 year: 1.97 and 0.44 years, and the phase of the planet is such that the minimum (or maximum) of the injected orbit is repeatedly hidden in the Solar Keepout gaps. These extreme misses represent $\lesssim$2\% of injected planets, and show that additional care needs to be taken when analyzing cases where the best-fitting orbit shows a minimum or maximum that is rarely sampled.}
\label{fig:period_harmonics_orbit_tracks}
\end{figure*}

Figure~\ref{fig:period_harmonics_orbit_tracks} shows the fits to these two period outliers. In both cases the minimum (or maximum) of the curve is repeatedly hidden by the Keepout gap, resulting in a fit that misses the period by either a factor of 2 or 3. For actual on-sky data, it will be necessary to exercise additional caution for recovered orbits such as in Figure~\ref{fig:period_harmonics_orbit_tracks}, where the maximum and/or minimum of the curve is repeatedly hidden in the gaps. Our tests indicate this is a relatively rare outcome, with only 2 out of 111 cases showing these significant period misses.

\begin{figure*}[t]
    \centering
    \includegraphics[width=\textwidth]{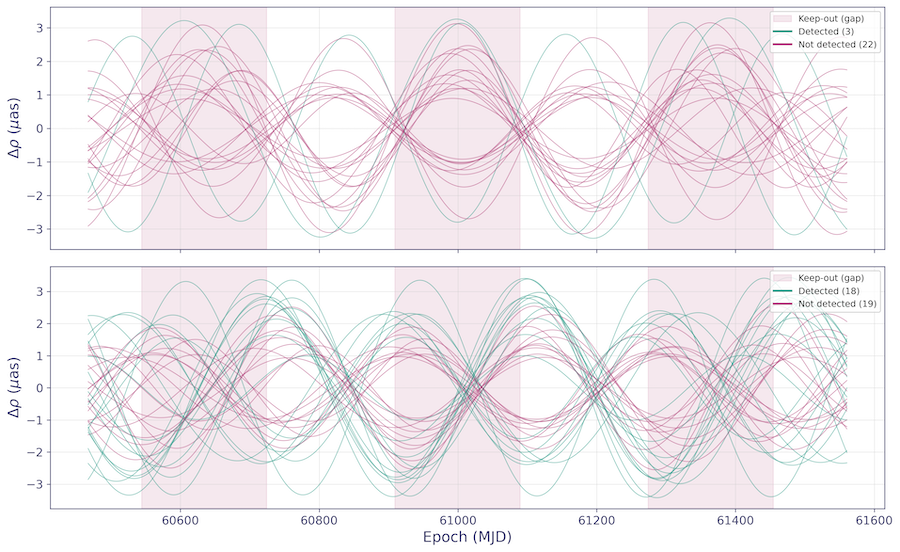}
    \caption{The Solar Keepout (red shading) for the SHERA mission results in $\sim$6 month gaps when a target system cannot be observed, resulting in decreased sensitivity to $\sim$1 year period planets. The top panel shows the astrometric signal of simulated planets with periods near 1 year and orbital phases such that the maximum of the curve (or the minimum) is continuously hidden by the gaps. These planets are recovered at a much lower rate (12\%) than planets in the lower panel, where both minimum and maximum of the signal are observed (49\%).}
    \label{fig:obsgap_phase_tracks}
\end{figure*}

We further investigate the impact of this effect by considering a series of simulated Level 3 datasets with orbital periods close to 1 year. We find that orbital phase with respect to the timing of the gaps has a significant impact on whether or not a planet is detected (based on the periodogram threshold), as shown in Figure~\ref{fig:obsgap_phase_tracks}. When both minimum and maximum of the astrometric signal are visible between the Keepout gaps, that planet is more likely to be detected (18 out of 37 cases). On the other hand, when the minimum (or maximum) is repeatedly hidden by the Keepout gaps, the planet is much less likely to be detected (3 out of 25 cases). While we will address fitting Level 1 data in a future paper, we expect binary relative astrometry to be even less sensitive to some 1 year orbits compared to Level 2 and Level 3 data, if the period and phase of the orbit matches the differential parallax signal as seen from the spacecraft.

\subsection{Fitting Datasets with More than One Planet}

\begin{figure*}[t]
    \centering
    \includegraphics[width=\textwidth]{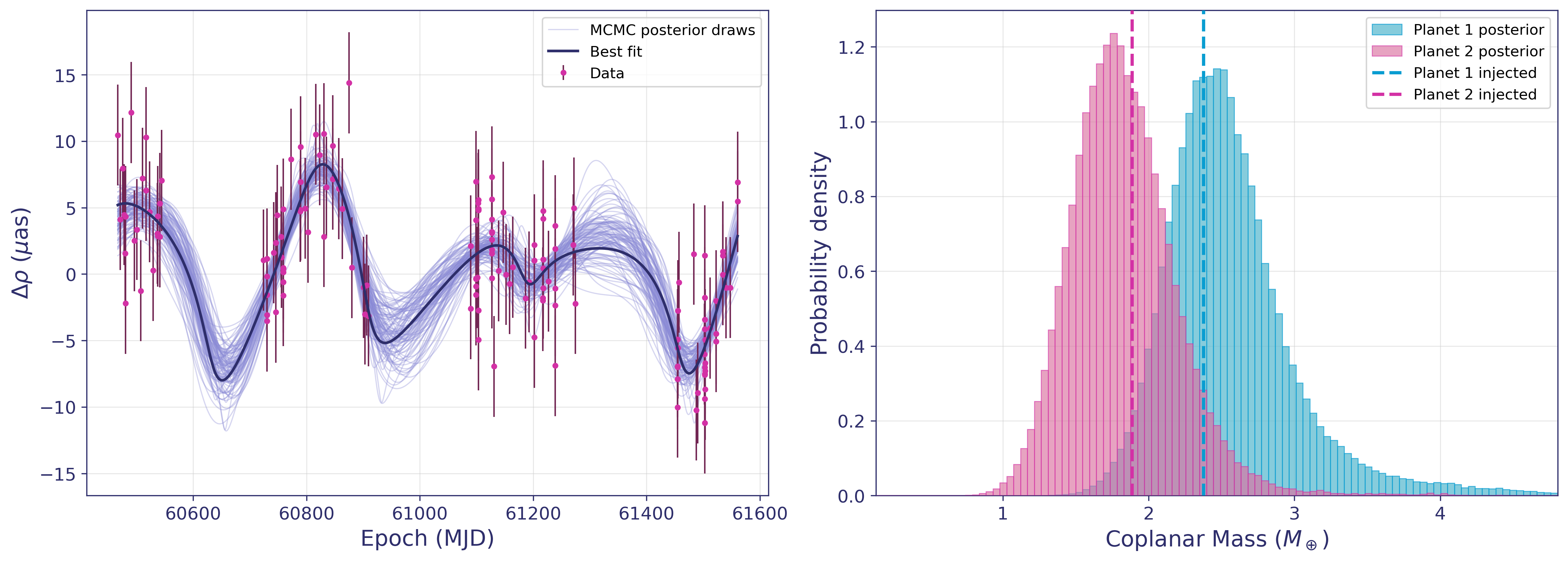}
    \caption{An example of a two-planet simulated SHERA dataset around $\alpha$ Cen A, and the MARA MCMC fit (left). The two planets, with periods of 0.754 and 1.069 yrs, are visible as overlapping periodic signals. On the right are the coplanar mass posteriors for each planet, which recover the injected values to within 1$\sigma$. With MARA, we are able to recover systems with multiple planets in addition to single-planet systems.}
\label{fig:two_planet_fit_and_posteriors}
\end{figure*}

In addition to the 10,000 simulated Level 3 datasets with only 1 injected planet, we also fit an additional 10,000 datasets with 2 injected planets, testing our ability to recover more than one planet in the same binary system with MARA. In many cases we are able to recover both planets, when both planets have mass and period similar to detectable planets in 1-planet datasets. Figure~\ref{fig:two_planet_fit_and_posteriors} shows an example of a two-planet dataset and the results from the MCMC fit. The period and coplanar mass of both planets are recovered to within 1$\sigma$ of the injected value.

There is a small amount of decreased sensitivity to planets in two-planet datasets compared to single planet ones. For example, for injected planets between 1.85 and 2.05 years, and masses between 0.9 and 1.1 M$_\oplus$, we recover 87\% of planets in single planet datasets; this drops to 79\% in two-planet datasets. Most of these lost planets have a similar period to a more massive planet in the data, or otherwise have a combination of periods and phases such that the clearest deviations from a 1-planet fit fall in the observing gaps.

\subsection{Fits to Level 2 and Level 3 Data}

Fits and analysis presented above are based on Level 3 Data, where any changes in separation due to to the binary orbit and system space motion have been removed. We also perform MCMC fits on Level 2 data for 12 systems, and compare the posteriors to the Level 3 fits. We find general overlap between the two sets of posteriors. This is not surprising, since there is generally very little covariance between orbital parameters of the planet and of the binary stars. The binary orbital parameters are overall very well constrained, and in general we would not expect uncertainties in a $\gtrsim$100 yr orbit to appear as periodic signals with a $\sim$1 year period.

\begin{figure*}[t]
    \centering
    \includegraphics[width=\textwidth]{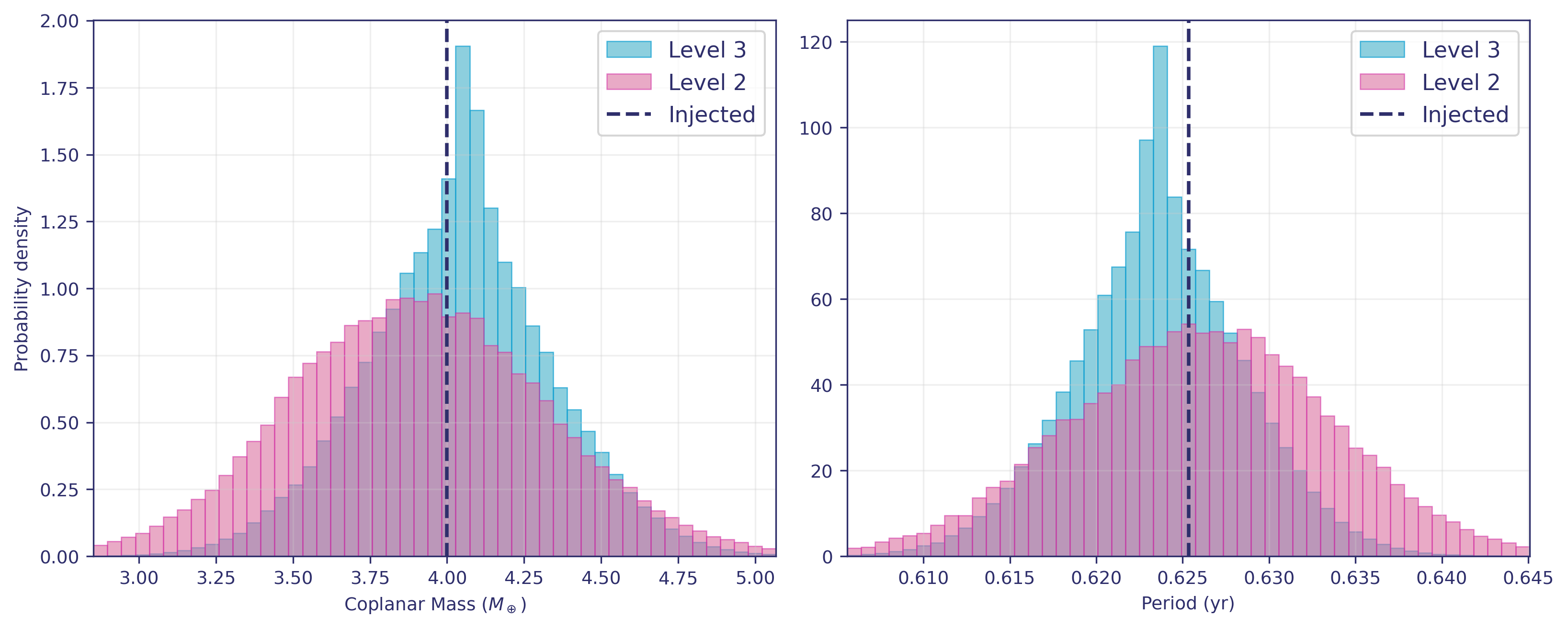}
    \caption{A comparison of the posteriors from MCMC fits to the same simulated planet with Level 2 data (pink) and Level 3 data (blue), for both coplanar mass (left) and period (right). Despite the longer computation time required to fit Level 2 data, the two posteriors overlap with slightly less precision on Level 2 fits. The two posteriors above recover the planet to be $4.0 \pm 0.3$ M$_\oplus$ and $0.624 \pm 0.005$ yrs in Level 3 data, and $3.9 \pm 0.4$ M$_\oplus$ and $0.626 \pm 0.007$ yrs in Level 2 data. This overlap allows us to use Level 3 fits as a proxy for fits to full SHERA data when computing sensitivity.}
    \label{fig:mode_comparison_posteriors}
\end{figure*}

Figure~\ref{fig:mode_comparison_posteriors} shows this comparison for one representative case. The injected planet has a coplanar mass  of 4.0 M$_\oplus$ and period of 0.625 yrs. The recovered posterior is $4.0 \pm 0.3$ M$_\oplus$ and $0.624 \pm 0.005$ yrs for the fit to Level 3 data, compared to $3.9 \pm 0.4$ M$_\oplus$ and $0.626 \pm 0.007$ yrs for Level 2 data. The main systematic difference seen in these comparisons is that posteriors from Level 2 fits are shifted to slightly lower coplanar planet masses and slightly larger period. The posteriors are expanded as well, by about 40\% in coplanar mass and period. The likely cause of these changes is the extent of the curvature in the binary star orbit not being fully constrained by the priors on the binary orbit parameters. As a result, some of the planet signal in the separation data is absorbed into the binary star's orbital parameters. While the planet parameters in Level 2 fits are less constrained compared to Level 3 fits, these tests confirm that it is reasonable to use Level 3 fits as a baseline when computing completeness to planets.

Level 2 data also allows us to examine a potential science return of a binary relative astrometry mission: how 1D astrometry at the microarcsecond level can improve our knowledge of the orbital parameters of some of the closest Sun-like stars in binary systems. Preliminary results for $\alpha$ Cen with simulated Level 2 SHERA data show that there is an improvement in the posteriors on binary orbital parameters compared to the priors, where the priors are based on the high precision fit of \citet{akeson:2021}. For example, semi-major axis and period  precision on the $\alpha$ Cen AB orbit are expected to improve from $17.493" \pm 0.010"$ and $79.762 \pm 0.019$ yrs \citep{akeson:2021} to $17.494" \pm 0.003"$ and $79.760 \pm 0.011$ yrs by including 3 years of SHERA data. Parallax precision should also improve significantly from SHERA observations (following a fit to the differential parallax in Level 1 data), at which point we expect precision on the mass of $\alpha$ Cen A to improve from 0.0025 M$_\odot$ to 0.007 M$_\odot$. We expect even better constraints for the other systems on the SHERA target list, as their orbits are generally not as well constrained as $\alpha$ Cen.

Ultimately, actual binary relative astrometry observations will be fit at the Level 1 level. While we leave the development of a Level 1 version of the MARA pipeline for future work, we expect the major difference between Level 1 and Level 2 fits to be a decreased sensitivity to planets with orbital periods close to 1 year, since both differential parallax and planet signal must be fit simultaneously. This will impact planets with a phase similar to the expected parallax signal the most.

Our tests here show that Level 3 fits are a suitable proxy for the more computationally expensive Level 2 fits. Despite the slightly larger posteriors for Level 2 fits, the more efficient Level 3 fits allow us to more quickly determine what simulated planets can be recovered, and the typical accuracy of the recovered parameters.

\section{Sensitivity Analysis and Discussion}\label{sec:diss}

Following the validation of the MARA pipeline, we proceed to demonstrate how MARA can be used to compute the overall sensitivity of binary relative astrometry observations of a single host star, as a function of orbital period and planet mass. 

\subsection{Analytic Sensitivity Prediction}

The sensitivity of an astrometric planet search can be estimated from basic planet and observing properties. In particular, if we require a FAP of less than $10^{-3}$ (which corresponds to a 4$\sigma$ detection of a planet), the corresponding signal (the semi-major axis of the star's reflex orbit) must be four times larger than the astrometric noise at the end of the mission. The signal becomes a function of planet mass, stellar host mass, distance to the system, and planet semi-major axis. Final noise can be approximated as $\sigma_{Final} = \frac{\sigma_i}{\sqrt{N}}$, where $\sigma_i$  is the assumed single-epoch error ($3.8$ \uas\ for the SHERA mission concept), and N is the number of observations of that system. Finally, a correction factor is needed to account for partial orbits (for example, if only 3 years of data are taken on a 30 year orbit). We define the analytic prediction for binary relative astrometry sensitivity, then, as:

\begin{equation}
\frac{m_p}{M_\oplus} \gtrsim 
\left(\frac{4 \sigma_{Final}}{3.00\,\mu {\rm as}}\right) 
\left(\frac{M_*}{M_\odot}\right) 
\left(\frac{d}{1\,{\rm pc}}\right) 
\left(\frac{a_p}{1\,{\rm au}} \right)^{-1}
\label{eq:analytic_prediction1}
\end{equation}

\noindent if the orbital period $P < T_{baseline}$, the baseline of the observations in years. Otherwise:

\begin{equation}
\frac{m_p}{M_\oplus} 
\gtrsim 
\frac{4 \sigma_{Final}}{3.00\,\mu {\rm as}} 
\left(\frac{M_*}{M_\odot}\right)
\left(\frac{d}{1\,{\rm pc}}\right)
\left(\frac{a_p}{1\,{\rm au}}\right)^{-1}
\left(\frac{P}{T_{baseline}}\right)^2 
= \frac{4\,\sigma_{Final}}{3.00\,\mu {\rm as}} 
\left(\frac{d}{1\,{\rm pc}}\right) 
\left(\frac{a_p}{1\,{\rm au}} \right)^2 {T^{-2}_{baseline}}
\label{eq:analytic_prediction2}
\end{equation}

\noindent In these equations, $m_p$ is the smallest detectable planet mass and $M_*$ is the mass of the host star; we assume the planet mass is much smaller than the stellar host mass. $a_p$ is the planet's orbital semi-major axis, and $d$ is the distance to the host star. For astrometric observations, a lower-mass planet becomes more detectable generally around lower-mass, more nearby stars. Astrometry is more sensitive to wider-separation planets (since the size of the star's reflex orbit increases) out to an orbital period equal to the time baseline of the observations, with sensitivity decreasing at even longer periods.

\begin{figure*}[t]
    \centering
    \includegraphics[width=\textwidth]{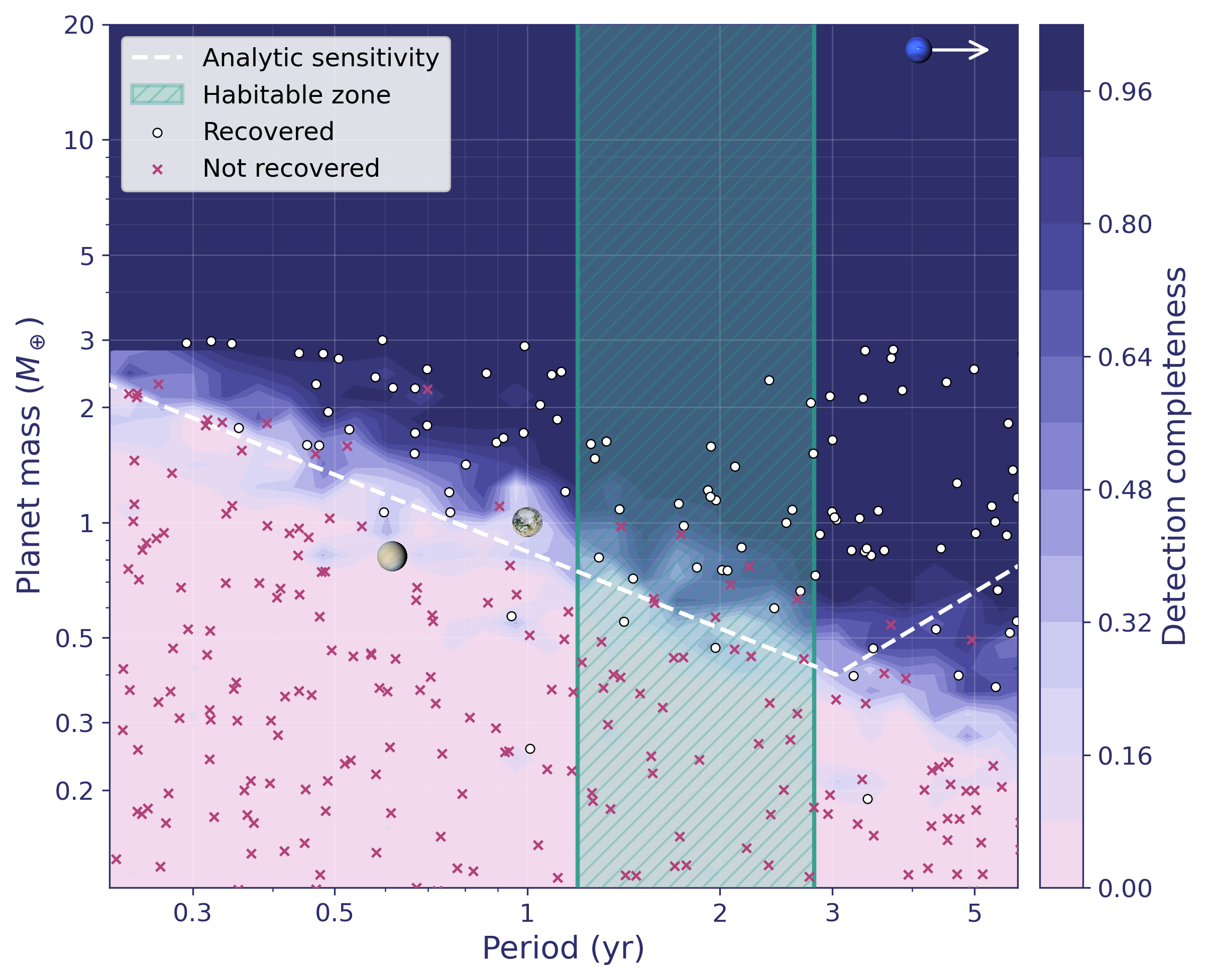}
    \caption{Completeness to planets for $\alpha$ Cen A, assuming 3 years of SHERA observations with 3.8\,\uas\ uncertainties with 149 total epochs. The white dashed line shows the analytic prediction (Equations~\ref{eq:analytic_prediction1} and \ref{eq:analytic_prediction2}), with the symbols showing Venus, Earth, and Neptune for comparison. Shaded contours give the results from 10,000 injection/recovery tests using the periodogram portion of the MARA pipeline, while circles and crosses are from a random subset of 400 injections. The shaded green region corresponds to the conservative habitable zone following \citep{Kopparapu2013}. There is generally good agreement between the two methods, with deviations largely due to combinations of orbital phase and the Solar Keepout gaps either making a planet more or less detectable than predicted by the analytic approximation. From this analysis SHERA observations would be about 87\% complete to a planet with mass between 0.9 and 1.1 M$_\oplus$ and period between 1.85-2.05 years, corresponding to the center of the habitable zone for $\alpha$ Cen A.  }
    \label{fig:tongue_plot_1planet}
\end{figure*}

The dashed line in Figure~\ref{fig:tongue_plot_1planet} shows the analytic prediction for the example of 3 years of SHERA observations of the $\alpha$ Cen system, for planets orbiting $\alpha$ Cen A. This prediction reaches masses below 1 M$_\oplus$ for periods from 0.8 to 6 years. This simplified analytic treatment provides a starting point for our sensitivity and completeness analysis, but it does not fully capture the complexity of recovering planets from 1D astrometry. In particular, the covariance between orbital phase, period, eccentricity, and the location of the Solar Keepout gaps is not part of this approximation. To fully account for these effects, we turn to injection/recovery tests with the MARA pipeline.

\subsection{Binary Relative Astrometry Sensitivity: Injection/Recovery Tests}

For a more robust measurement of the sensitivity to planets from binary relative astrometry we examine the results from the MARA pipeline to the 10,000 simulated SHERA Level 3 datasets. These datasets have planet masses between 0.05 and 3 M$_\oplus$ and periods from 0.2 to 6.5 years, and planets are considered recovered in the dataset if the periodogram FAP is less than $10^{-3}$. Results from these injection/recovery tests are shown in Figure~\ref{fig:tongue_plot_1planet} as shaded contours based on the fraction of injected planets that are recovered in each cell. These injection/recovery tests generally match the analytic prediction described above, though for orbital periods less than 3 years the transition from not recovered to recovered is more gradual for the periodogram analysis than it is for the sharp boundary of the analytic expression.

A large part of this change in sensitivity is due to the Solar Keepout, as discussed above in Section~\ref{sec:keepout}. Two planets with similar period and mass will be more detectable if the maximum and minimum of the astrometric signal is visible, or less detectable if either the maximum or the minimum takes place in the middle of a Keepout gap. In particular, orbital periods at 0.5, 1, and 2 years show less sensitivity to lower-mass planets compared to other periods, when the orbital period is equal to the one-year timing of the Keepout gaps, or harmonic of it. Nevertheless, the completeness to planets around $\alpha$ Cen A in the 1 planet case is excellent, with $\ge$87\% completeness for planets between 0.9-1.1 M$_\oplus$ in the center of the habitable zone ($1.85 yr \le P \le 2.05 yr$). At 1 year periods, these observations are 73\% complete to planets between 0.85 and 3 M$_\oplus$.

\begin{figure*}[t]
    \centering
    \includegraphics[width=\textwidth]{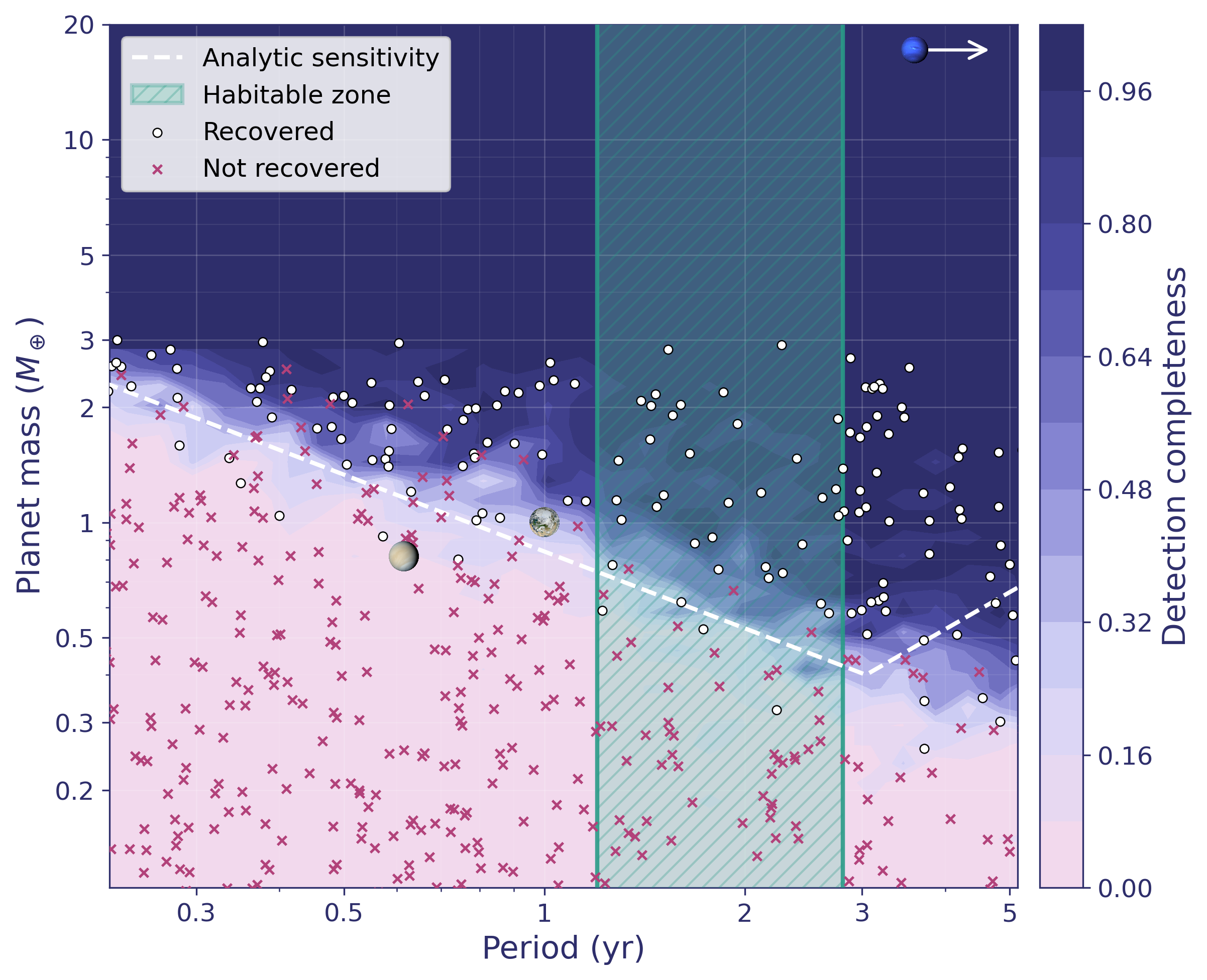}
    \caption{Completeness to planets for simulated Level 3 data for datasets with 2 planets each. Overall, sensitivity is similar to the 1 planet version in Figure~\ref{fig:tongue_plot_1planet}, but slightly reduced due to the added difficulty of finding a second, lower-mass planet in a dataset that already has a planet.}
    \label{fig:tongue_plot_2planet}
\end{figure*}

We expand this injection/recovery analysis to 2 planet systems in Figure~\ref{fig:tongue_plot_2planet}. The analytic prediction is the same as before, but datasets for both the periodogram analysis and MCMC are now the 2 planet versions, including 10,000 datasets for the periodogram analysis (contours), and 400 randomly drawn injected planets (circles and crosses). As expected based on our earlier analysis, SHERA sensitivity drops slightly when considering 2 planet systems. For example, while in the 1 planet cases our completeness to 0.9-1.1 M$_\oplus$ between 1.85 and 2.05 year periods is 87\%, this drops to 79\% for the 2-planet case.

Overall, we reach the same basic conclusion even with multiple planet systems and with increasingly detailed fits: using the MARA pipeline we are able to recover rocky planets from simulated SHERA data in the habitable zone. In the case of $\alpha$ Cen A, that corresponds to $\sim$1 M$_\oplus$ planets at a period of 1 year, and down to $\sim$0.5 M$_\oplus$ planets in 3 year orbits. The depth of search plots (or "tongue plots") presented here serve as the backbone for demographics analysis from binary relative astrometry missions, as both per-star completeness as a function of planet parameters and a list of detections are needed to solve for the underlying occurrence rate (e.g. \citealt{fulton:2021, nielsen:2019}). The methodology we lay out here will allow binary relative astromery missions to carry out one of the key science goals of such a mission: to measure the occurrence rate of planets in binaries, and determine if it is suppressed compared to single stars.

\section{Conclusions and Future Work}\label{sec:conclusion}
We have presented the MARA planet detection and characterization pipeline for binary relative astrometry. MARA is specifically designed for the unique format of these data, and directly addresses the perspective effects and speed of light issues needed to fit astrometry at the microarcsecond level. Using this pipeline we confirm that a mission using the binary relative astrometry technique with an individual measurement precision $\sim$4 \uas\ will be capable of detecting rocky exoplanets in the habitable zones of nearby binaries. Specifically, we have assessed example simulated data from the SHERA mission to compute sensitivity to $\alpha$ Centauri A, for which MARA can detect planets down to $\sim 0.5$ $M_\oplus$. We have also confirmed that we are able to disentangle multiple planets from a single dataset, detect planets near $1$ year periods, and precisely measure period even with realistic observing cadences and gaps due to the solar exclusion zone. Our injection/recovery tests broadly agree with our analytical predictions for binary relative astrometry performance.

Going forward, we aim to expand these results to other nearby binaries, including the SHERA target list \citep{christiansen:2026}, with simulated datasets for more binaries than just $\alpha$ Cen. Given Equation~\ref{eq:analytic_prediction1}, we expect mass sensitivity at a given orbital period to scale as $m_p \propto M_* d$ (with an additional instrument performance term based on the apparent magnitude of the star). As a result, while SHERA sensitivity to 1 year period planets for the 1.3 pc $\alpha$ Cen A is $\sim$0.5 M$_\oplus$, it should rise to $\sim$6 M$_\oplus$ for the most distant SHERA target, HR 2667 at 17.1 pc. Verifying these results with injection/recovery tests will help determine how accurate these predictions and scaling relationships are. In addition, expanding to the full list will allow us to perform a detailed investigation into how large a role accurate knowledge of the binary orbit plays in planet sensitivity. While $\alpha$ Cen has the best constrained binary orbit of the full target list \citep{akeson:2021}, it is likely that the increased distance to the other systems (and their generally larger orbital periods) means a smaller amount of binary orbit precision has less of an impact on planet parameters, but this too will be studied. This expanded analysis will be particularly useful for planning future binary relative astrometry missions that target more distant binaries.

We are also working to improve MARA so that in addition to fitting simulated Level 2 and Level 3 data, we can also fit Level 1 data, the same type of data a binary relative astrometry mission like SHERA or TOLIMAN will actually produce in flight. This will allow us to directly validate the use of Level 2 and Level 3 data as proxies for the actual mission dataset. At the same time we aim to fit more complex simulated datasets, examining the effect of different cadence and revisit strategies, and to simulate planets on eccentric orbits, in addition to circular orbits. This change in particular should provide better evidence for the origin of the small bias in recovered MCMC coplanar mass compared to injections and how to address it with improved priors in the fit.

Finally, we will also expand MARA to accept radial velocity measurements of either (or both) host stars in the binary, to help determine the 3D architecture of more massive planets in the system. This change in particular will help break the 1D degeneracy we discuss above, and directly measure the mass (rather than coplanar mass) of these planets, especially when coupled with Extreme Precision Radial Velocities (EPRVs). All of these changes together will enable MARA to be used for planning or analyzing data from the SHERA and TOLIMAN observations, as well as any future mission that utilizes binary relative astrometry as a planet search technique.

Binary relative astrometry missions will be sensitive to a class of rocky planets that have remained beyond the reach of modern detection techniques, reaching into the habitable zone of nearby Sun-like stars in binaries. With MARA we are able to identify and characterize planets in these data, as well as compute the completeness to planets needed for demographics calculations. Ultimately, missions like SHERA or TOLIMAN will play a valuable role in preparing for the Habitable Worlds Observatory, both in screening for planets around high-priority targets, but also determining how the occurrence rate of planets depends on binarity \citep{christiansen:2026}. As we show here, these binary relative astrometry missions will be able to detect and characterize an Earth analog around $\alpha$ Cen A, sensitive to small planets in the habitable zone of these binary targets like no planet finding technique could previously.

\section{Acknowledgments}

Part of this research conducted at NMSU was supported by subcontracts 1701901 and 1718858 from the Jet Propulsion Laboratory, California Institute of Technology, under a contract with the National Aeronautics and Space Administration (80NM0018D0004).
Part of this research was carried out at the Jet Propulsion Laboratory, California Institute of Technology, under a contract with the National Aeronautics and Space Administration (80NM0018D0004). 
W.R. and E.L.N acknowledge support from NSF AAG 2510959. 
K.M.K. acknowledges support from NSF AAG 2407522.
This work utilized resources from the New Mexico State University High Performance Computing Group 
\citep{Trecakov2021}, which is directly supported by the National Science Foundation (OAC-2019000), the Student Technology Advisory Committee, and New Mexico State University and benefits from inclusion in various grants (DoD ARO-W911NF1810454; NSF EPSCoR OIA-1757207; Partnership for the Advancement of Cancer Research, supported in part by NCI grants U54 CA132383 (NMSU)).
This research has made use of the NASA Exoplanet Archive, which is operated by the California Institute of Technology, under contract with the National Aeronautics and Space Administration under the Exoplanet Exploration Program.

\facilities{CHTC \citep{CHTC}}

\bibliography{refs}

\end{document}